\documentclass[manuscript]{aastex62}
\usepackage[encapsulated]{CJK}
\usepackage{mathrsfs}
\usepackage{amsmath}
\usepackage{longtable}
\usepackage{booktabs}
\usepackage{hyperref} 
\usepackage{overpic}
\def\etal {et al.~}

\newbox\grsign \setbox\grsign=\hbox{$>$} \newdimen\grdimen \grdimen=\ht\grsign
\newbox\laxbox \newbox\gaxbox
\setbox\gaxbox=\hbox{\raise.5ex\hbox{$>$}\llap
     {\lower.5ex\hbox{$\sim$}}}\ht1=\grdimen\dp1=0pt
\setbox\laxbox=\hbox{\raise.5ex\hbox{$<$}\llap
     {\lower.5ex\hbox{$\sim$}}}\ht2=\grdimen\dp2=0pt

\shorttitle{CO OUTFLOW SURVEY TOWARD W3/4/5}
\shortauthors{Li \etal}

\newcommand{\co}{$^{12}$CO }                             
\newcommand{\xco}{$^{13}$CO }                            

\definecolor{malachite}{rgb}{0.34, 0.7, 0.22}

\begin{document}

\title{CO Outflow Candidates Toward the W3/4/5 Complex I: The Sample and its Spatial Distribution}

\correspondingauthor{Ye Xu}
\email{xuye@pmo.ac.cn, liyj@pmo.ac.cn}

\author{Yingjie Li}\affiliation{Purple Mountain Observatory, Chinese Academy of Sciences, Nanjing 210008, China}
\affiliation{University of Science and Technology of China, Chinese Academy of Sciences, Hefei, Anhui 230026, China}

\author{Ye Xu}
\affiliation{Purple Mountain Observatory, Chinese Academy of Sciences, Nanjing 210008, China}

\author{Yan Sun}
\affiliation{Purple Mountain Observatory, Chinese Academy of Sciences, Nanjing 210008, China}

\author{Qing-Zeng Yan}
\affiliation{Purple Mountain Observatory, Chinese Academy of Sciences, Nanjing 210008, China}

\author{Yuehui Ma}
\affiliation{Purple Mountain Observatory, Chinese Academy of Sciences, Nanjing 210008, China}
\affiliation{University of the Chinese Academy of Sciences, 19A Yuquan Road, Shijingshan District, Beijing 100049, People's Republic of China}

\author{Ji Yang}
\affiliation{Purple Mountain Observatory, Chinese Academy of Sciences, Nanjing 210008, China}

\begin{abstract}
Using the Purple Mountain Observatory Delingha 13.7 m telescope, we conducted a large-scale $^{12}$CO $\left(J=1 \rightarrow 0\right)$ outflow survey (over $\sim$ 110 deg$^2$) toward the W3/4/5 complex and its surroundings. In total,  459 outflow candidates were identified. Approximately 62\% (284) were located in the Perseus arm, including W3/4/5 complex and its surroundings, while $\sim$ 35\% (162) were located in the Local arm, $\sim$ 1\% (5) in the Outer arm, and $\sim$ 2\% (8) in two interarm regions. This result indicated that star formation was concentrated in the Galactic spiral arms. The detailed spatial distribution of the outflow candidates showed that the Perseus arm presented the most active star formation among the study regions. The W3/4/5 complex is a great region to research massive star formation in a triggered environment. A key region, which has been well-studied by other researches, is in the eastern high-density W3 complex that neighbors the W4 complex. Two shell-like structures in the Local arm contain candidates that can be used to study the impact on star formation imposed by massive or intermediate-mass stars in relatively isolated systems. The majority of outflow candidates in the two interarm regions and the Outer arm are located in filamentary structures.
\end{abstract}


\section{Introduction}

Associated with all stages of early stellar evolution, from deeply embedded protostellar objects to optically visible young stars, outflows are ubiquitous in star-forming regions \citep{RB2001}. They are intrinsic processes that are related to the mass-loss phase of both low- and high-mass stars \citep[e.g.,][]{ASG2007}. Entraining and accelerating ambient gas, outflowing supersonic winds produce molecular outflows, which in turn affect the dynamics and structure of their  parent clouds \citep{NS1980, ABG2010}.  For moderate- and high-mass stars, such outflow activity is followed by ever more powerful momentum and energy injection mechanisms, such as ejection by UV radiation, ionizing radiation, stellar winds and even explosions \citep{FRC2014, B2016}. These processes may play a role in determining the star formation efficiency (SFE) in cluster environments \citep{E1998}, and perhaps sculpturing the shape of the stellar initial mass function \citep[e.g.,][]{AF1996, PBK2010}.

CO spectral lines are capable of revealing outflow activities in star forming regions  \citep[e.g.,][]{SAL1987}. In comparison to optical outflows, CO outflows occur during younger stages \citep{GBW2011}. Having velocity information, CO transitions can be used to roughly measure the mass and momentum ejected from protostars or swept by ejecta \citep{B1996, GBW2011}. Therefore, CO outflows are used to investigate star formation activities in this work.

Large-scale single dish surveys can provide superb samples that merit further investigations with higher resolution facilities such as interferometers. Due to their large beam sizes (typically larger than 5$\arcsec$), single dish surveys usually take several cluster outflows as a single one \citep{MSG2016, LLX2018}. This also happens with interferometers; for instance, a single outflow observed with several 0.1$\arcsec$ in the G31.41+0.31 \citep[using the VLA and SMA,][]{AHK2008, CBZ2011} was proved to be a double-jet system traced by water masers \citep{MLC2013} with angular resolutions $\sim 1.6$ mas $\times 1.0$ mas using the VLBA. However, a large beam yields a fast survey speed, and consequently, single dish surveys usually cover large sky regions, which provide relatively complete samples for further studies using  interferometers. For example, we could use single dishes to identify outflow candidates and investigate them subsequently with more powerful telescopes, which is the purpose of this work.


Giant HII regions such as the W3, W4 and W5 (W3/4/5) Complexs, which are excited by the Cas OB6 association of stars, are nearby massive star forming regions on the Perseus arm in the outer Galaxy \citep{W1958, HT1998}.  Because of its importance to high-mass star formation studies \citep{GBW2011}, the entire W3/4/5 complex has been investigated insensitively in multi-wavelength bands that cover the X-ray \citep[e.g.,][]{RFG2016}, optical \citep[e.g.,][]{SBC2017}, infrared \citep[e.g.,][]{RMP2011, RMP2013},  and (sub-)millimeter and centimeter \citep[e.g.,][]{WBG2003, FS2007, DAM2014}, and different tracers including outflows \citep[e.g.,][]{BMR2002, GBW2011, ZZW2011}, masers \citep[e.g.,][]{SSE2001, MXC2010}, HII regions \citep[e.g.,][]{HAH2012}, and so on. However, most outflow surveys either detected a small number of outflows \citep[e.g.,][]{BMR2002} or were confined to relatively small regions  \citep[e.g., the search for outflows only in the W5 complex by][]{GBW2011}. In addition, gas clouds in the W3/4/5 complex provide a great opportunity for studying the range of properties of outflow and star-formation in different spiral arms and in interarm regions. Therefore, it is still meaningful to conduct a larger-scale (unbiased) \co (1 $\rightarrow$ 0) (frequently used as an outflow tracer) outflow survey towards the W3/4/5 complex and its surroundings.

Sun et al. (2019, in preparation) is going to study the structures and physical properties of the molecular (\co and its other two isotopic molecules) gas toward the W3/4/5 complex and its surroundings, which include the Local arm, the Perseus arm (the W3/4/5 complex and other clouds/complexes), the Outer arm and the gaps between spiral arms (see the summaries of the distances and velocity ranges of the entire observed area in Table \ref{tab:distance}).

\begin{deluxetable}{lccl}
\centering
\setlength\tabcolsep{15pt}
\tablecolumns{4}
\tabletypesize{\normalsize}
\tablewidth{15cm}
\tablecaption{Summary of Distances and Velocity Ranges for $^{13}$CO\label{tab:distance}}
\tablehead{
 \colhead{Sub-region} & \colhead{Distance} & \colhead{Velocity Range} & \colhead{Reference} \\
 \colhead{}   & \colhead{(pc)} & \colhead{(km s$^{-1}$)} & \colhead{}
}
\startdata
Local arm         & 600  & [-20, 7)    & 1 \\
Interarm 1        & 1280 & [-30, -20)  & 1 \\
Perseus arm       & 1960 & [-62, -30)  & 2, 3, 4\\
Interarm 2        & 3975 & [-68, -62)  & 1 \\
Outer arm         & 5990 & [-88, -68]  & 2, 3 \\
\enddata
\tablewidth{15cm}
\tablerefs{(1) Sun et al. in preparation, and references therein; (2) \citet{RMZ2009}; (3) \citet{RMB2014}; (4) \citet{XRZ2006}.}
\end{deluxetable}

The remainder of the paper is organized as follows. In Section 2, we describe the data used in this work. Next, in Section 3 we present the outflow detection process and our estimation of, including their statistics, their physical parameters, and compare the identified outflow candidates with the CO (3 $\rightarrow$ 2) outflow survey of \citet{GBW2011}. In Section 4, we describe the spatial distribution of the outflow candidates, which are followed by discussions about star formation activities and triggered star-formation in Section 5. A summary of the main results is given in Section 6.

\section{Data}

The Milky Way Imaging Scroll Painting (MWISP) project led by the Purple Mountain Observatory (PMO) is a large ongoing project whose goal is to map CO and its isotopic transitions towards the Galactic plane \citep{SYZ2018}. We selected a pilot region of $\sim$ 110 deg$^2$ ($129\degr.75\leq l\leq140\degr.25$, $-5\degr.25\leq b\leq5\degr.25$) to search for outflows by using the data of \co (J = 1 $\rightarrow$ 0) (115.271 GHz) and \xco (J = 1 $\rightarrow$ 0) (110.201 GHz).\footnote{All MWISP data cubes (including that used in this work) that have been used for publications are accessible now by contacting \url{dlhproposal@pmo.ac.cn}.} The $^{12}$CO and \xco molecular lines were observed from November 2011 to November 2017 using the Purple Mountain Observatory Delingha (PMODLH) 13.7 m telescope with the nine-beam superconducting array receiver (SSAR). SSAR works in the sideband separation mode and uses a fast Fourier transform spectrometer \citep{ZLS2011, SYS2012}.

The spectral resolution was 61 kHz, which is equivalent to a velocity resolution of $\sim0.16$ km s$^{-1}$ for \co and $\sim0.17$ km s$^{-1}$ for $^{13}$CO. The half-power beam-width (HPBW) was 49$\arcsec$ for $^{12}$CO, and 51$\arcsec$ for $^{13}$CO, and the data were gridded to 30$\arcsec$ pixels for both transitions. During the observations, the typical system temperature was $\sim280$ K for $^{12}$CO, and $\sim185$ K for $^{13}$CO. The main beam root mean squared noise (RMS) after main beam efficiency correction for a single 61 kHz channel was $\sim0.45$ K for $^{12}$CO, and $\sim0.25$ K for $^{13}$CO.

In the process of data reduction, the baseline was fitted with a first order (or linear) profile for the CO spectra. The baseline fluctuation which was evaluated by the RMS of the mean spectrum over $10\arcmin \times 10\arcmin$ was $\lesssim 0.13$ K for \co and $\lesssim 0.07$ K for $^{13}$CO.

\section{Data Analysis and Result}

\subsection{Outflow Identification}

A set of semi-automated IDL scripts \citep[for details see][]{LLX2018}\footnote{\url{https://github.com/liyj09/outflow-survey-v1}.} was used to search for outflows. Specifically, the scripts are based on longitude-latitude-velocity space, which is used to trace the cores found in the three-dimensional \xco data, and then search for and identify outflows in the three-dimensional \co data. After searching for \co velocity bulges\footnote{CUPID \citep[][part of the STARLINK software]{BRJ2007} clump-finding algorithm FELLWALKER \citep{B2015} were used (see \url{http://starlink.eao.hawaii.edu/starlink}).} based on the \xco peak velocity distribution maps, and after conducting line diagnoses of the positions with velocity bulges (which require the outflow candidate's position and velocity range), the outflow candidate sample was finally obtained \citep[see][]{LLX2018}.

During this process, we also built a scoring system to estimate the quality of the outflow candidates based on their line profiles, contour morphologies, and their P-V (Position-Velocity) diagrams \citep[see the detailed criteria in table 10 of ][]{LLX2018}. The scores range  from ``A'' to ``D'', where ``A'' denotes the most reliable outflow candidates, which are typically high-velocity ones and ``B'' denotes reliable outflow candidates (which include the majority of outflow candidates). The characteristics of the outflows are not obvious for those with quality level ``C'', but we cannot rule out the possibility of outflow activities in those candidates. Outflow candidates with score ``D'' were removed from our candidate  samples, owing to their ambiguous characteristics. For more detail of the quality levels (scores) see \citet{LLX2018}.

If the blue and red lobes were close (within 1.5 pc) or showed similar core component structures, they were paired to form a bipolar outflow candidate \citep[see][]{LLX2018}. The score and quality level of a bipolar outflow candidate was determined by the highest level of the two lobes. Mis-assignations are inevitable owing to the moderate spatial resolution of the data considered in this work (i.e., the HPBW is 49$\arcsec$ for $^{12}$CO). To avoid the negative impact of these flaws, we treated outflow candidates as single candidate lobes when we compared outflow candidates to each other.

\begin{figure}[!ht]
\centering
\includegraphics[height=0.38\textheight,angle=-90]{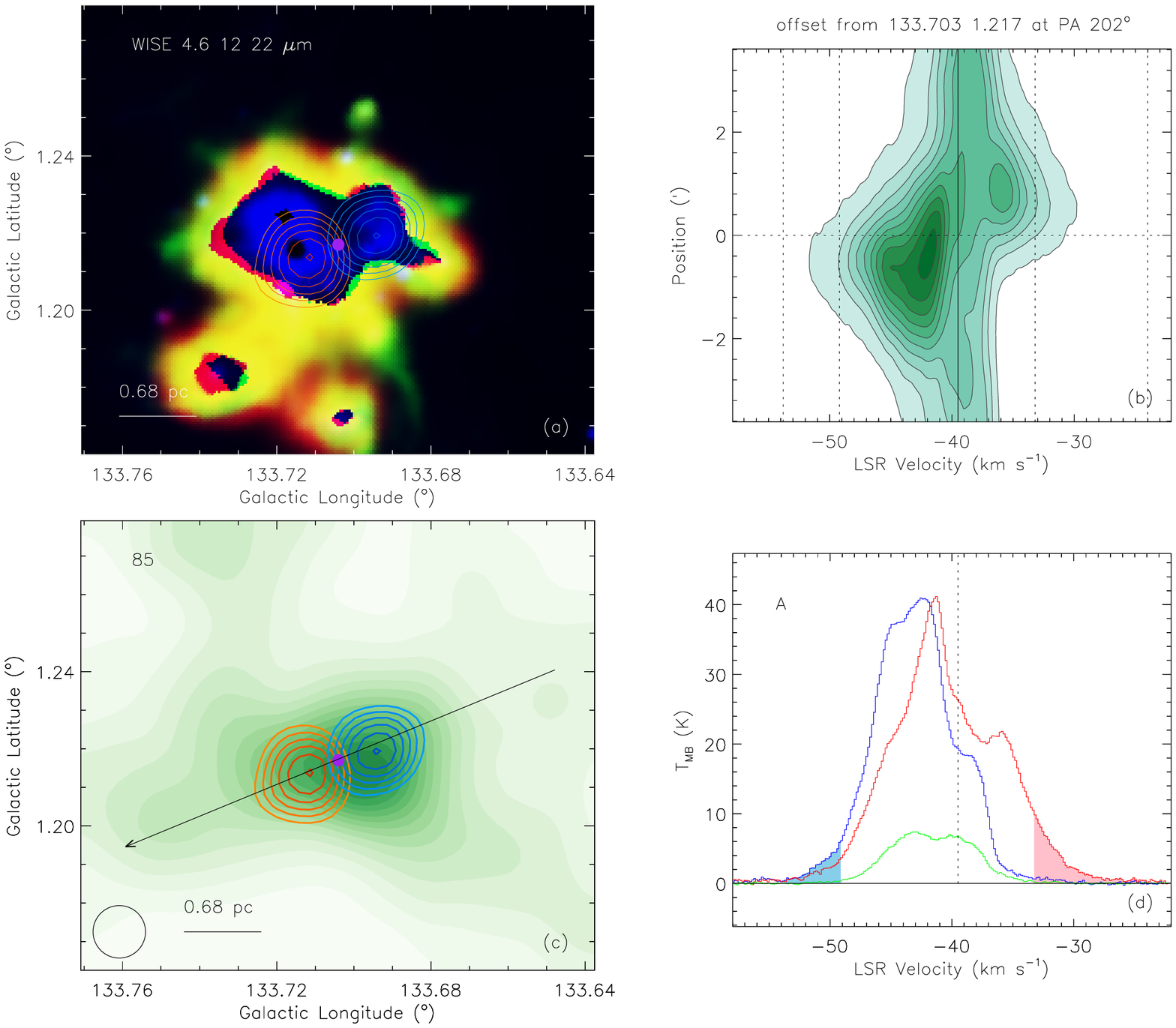}
\vrule height -5cm width 0.1pt
\includegraphics[height=0.38\textheight,angle=-90]{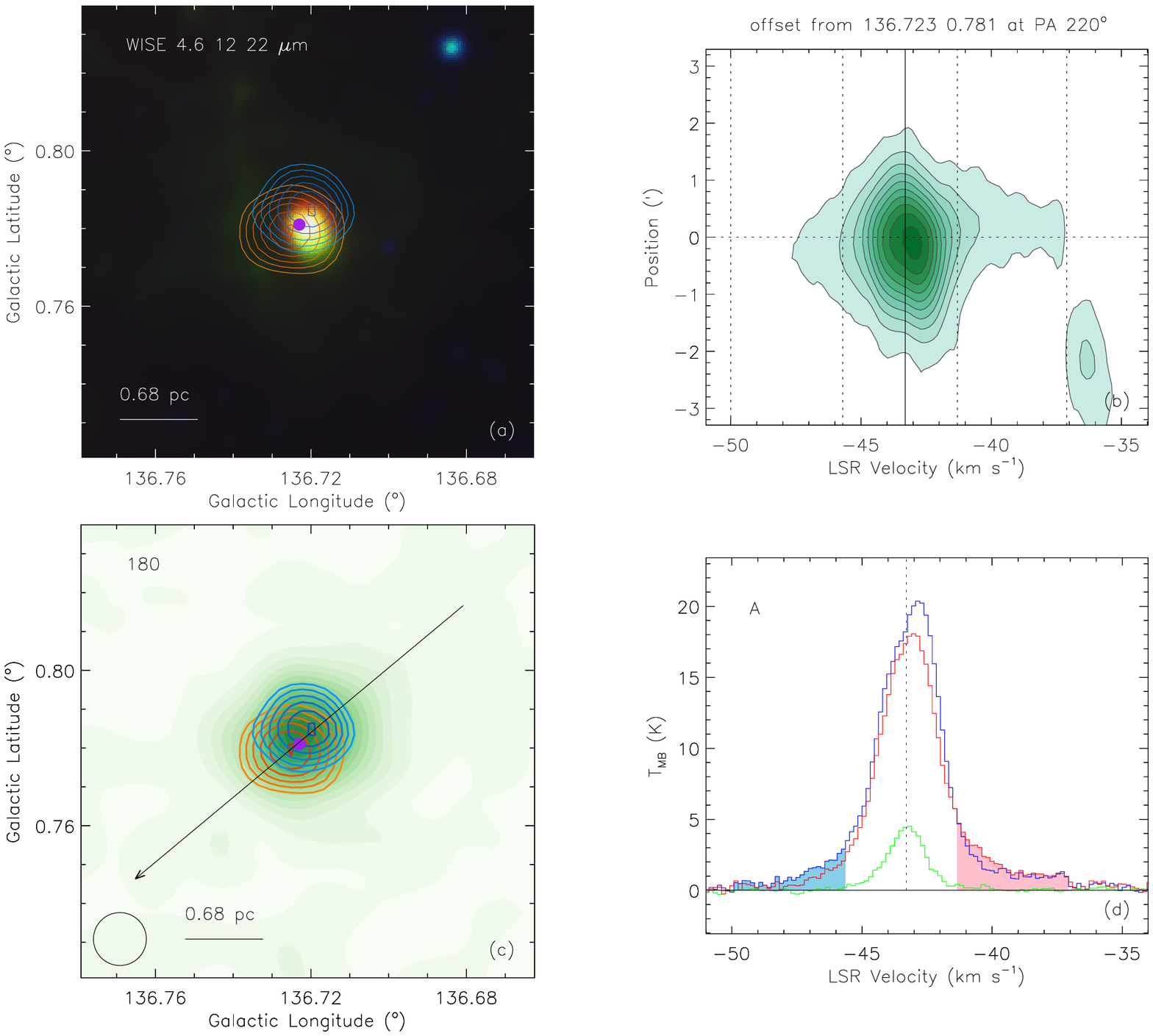}
\rule[-2pt]{3cm}{0.1pt}
\\[-0.5cm]
\includegraphics[height=0.38\textheight,angle=-90]{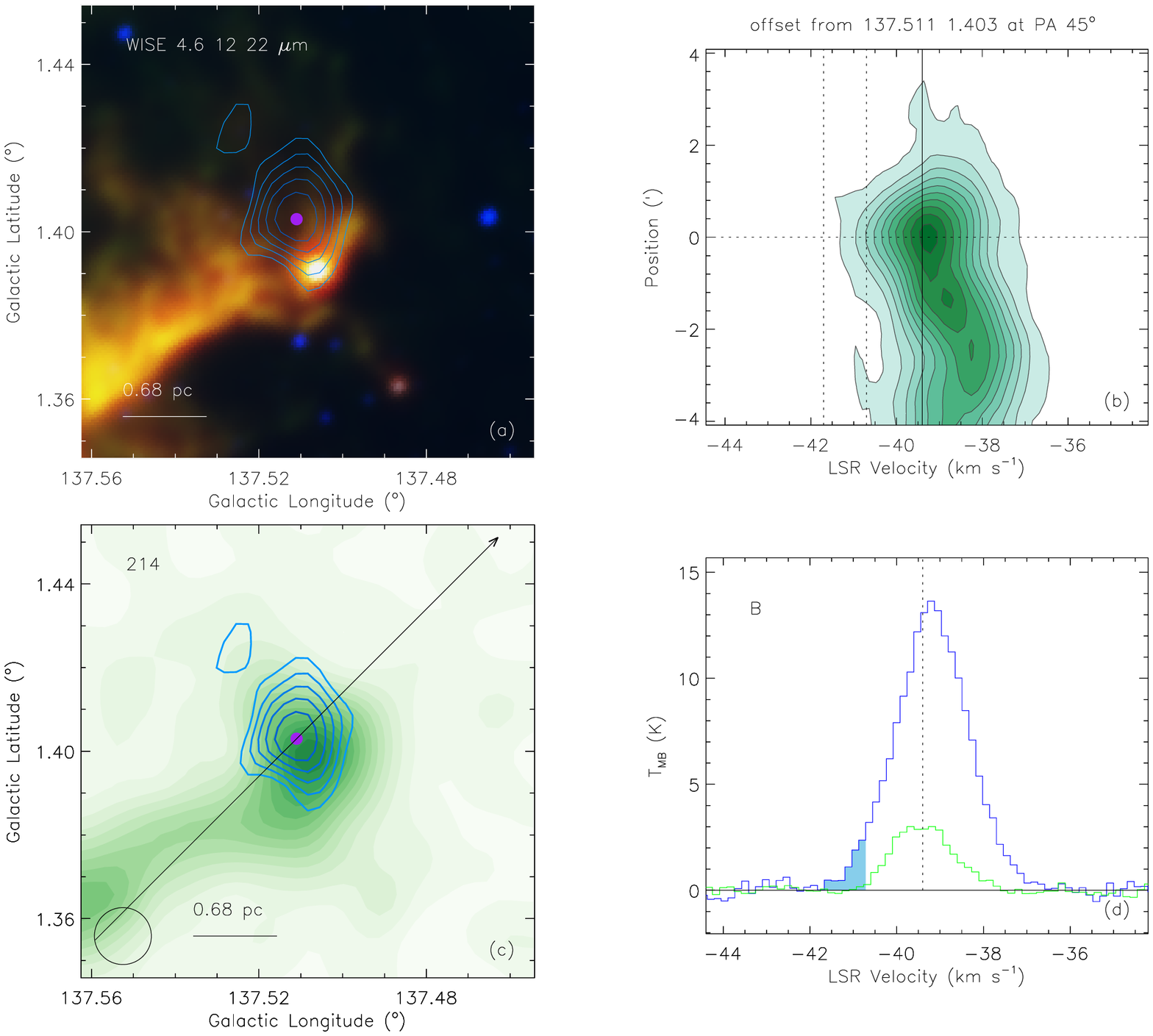}
\vrule depth 2.5cm width 0.1pt
\includegraphics[height=0.38\textheight,angle=-90]{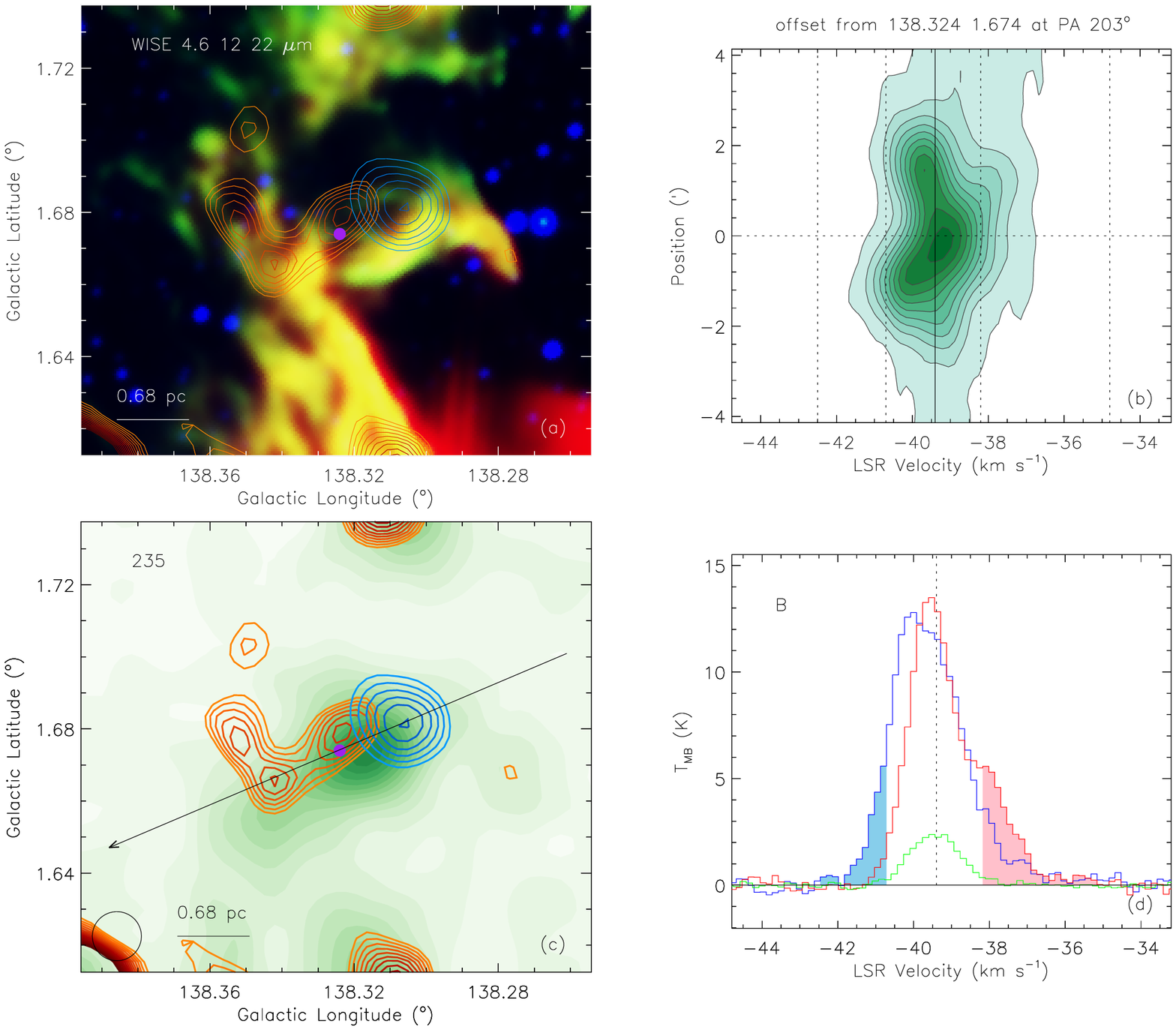}
\caption{Four example outflow candidates with indexes 85, 180, 214, 235 (see the indexes in panel (c) for each). (a): blue/red lobe contours superpose on WISE false color (blue, green and red for 4.6, 12 and 22 $\mu$m, respectively) image. (b): P-V diagram is along the black arrow (the position angle, PA, is reported at the top of the panel) which is shown in panel (c); the width to draw the P-V diagram is 1$\arcmin$ (2 pixels). (c): blue/red \co lobe contours overlay on the integrated intensity of \xco cores depicted as green-filled contours. (d): the blue/red spectrum is the \co averaged over $3\times3$ pixels centered at the blue/red emission peak positions, respectively; the green spectrum is the \xco averaged over $3\times3$ pixels centered at the purple point (denotes the position of outflow candidate) in panel (a) or (c); the blue/red line wing velocity range of each outflow candidate is specified with shading; the black dotted line in the center is of the peak velocity of $^{13}$CO. The physical scale or HPBW of $^{12}$CO is shown in the left-bottom corner of panels (a) or (c). The quality level of each outflow candidate is reported in the top left corner of panel (d). \\
(The complete figure set (459 images) is available in the online material.)}
\label{Fig:samples}
\end{figure}

Wide-field Infrared Survey Explorer (WISE) sources, especially those with excesses in the 12 $\mu$m and 22 $\mu$m bands, are indicators of star formation activity \citep{WEM2010}. WISE images were therefore useful to search for the driving sources of the outflow candidates. We superimposed the contours of an outflow candidate onto its corresponding WISE image and checked by eye for a point of light in the WISE image (e.g., see panel (a) in Figure \ref{Fig:samples}). These maps were used to judge whether the outflow candidates had any WISE associations \citep[for detail criteria see][]{LLX2018}. Future high-resolution investigations will be of great interest to confirm these preliminary associations.

We have displayed the outflow candidates in Figure \ref{Fig:samples}, while a list of the outflow candidates' positions and other properties are given in Table \ref{table:examples}, and a summary of the outflow survey results are provided in Table \ref{table:Statistics}. In Figure \ref{Fig:samples}, for a given bipolar outflow candidate, the integral velocity range of its corresponding \xco core component is indicated by the shading shown in panel (d). For a monopolar outflow candidate, the integral velocity ranges from the incipient velocity of the blue/red line wing to its symmetrical point with the symmetry axis of the peak velocity. The number of contour levels of the \xco integrated intensity was 15, and the contour intervals are 1/15 of the difference between the maximum and minimum integrated intensity in the mapped region. The incipient value of the contours is typically $\gtrsim 40$\% of the maximum integrated intensity in mapped regions.

\begin{deluxetable}{crrcccc}
\setlength\tabcolsep{10pt}
\tablecolumns{8}
\tabletypesize{\small}
\tablecaption{The Outflow Candidates \label{table:examples}}
\tablehead{
 \colhead{Index} & \colhead{$l$} & \colhead{$b$} & \colhead{Blue Line Wing} & \colhead{Red Line Wing} & \colhead{Quality level} & \colhead{WISE Detection} \\
 \colhead{} & \colhead{(\degr)} & \colhead{(\degr)} & \colhead{(km s$^{-1}$)} & \colhead{(km s$^{-1}$)} & \colhead{Blue$|$Red} & \colhead{}
}
\startdata
\multicolumn{7}{l}{the Perseus arm}\\
\hline
  1  & 130.136 & -4.467   & $(-44.3, -43.0)$  &      \nodata       &   B    &  ? \\
  2  & 130.155 & -4.537   & $(-44.6, -43.3)$  &      \nodata       &   B    &  Y \\
  3  & 130.399 & -0.728   &     \nodata       &  $(-29.8, -28.4)$  &   B    &  ? \\
  4  & 130.401 &  1.667   &     \nodata       &  $(-41.5, -40.4)$  &   C    &  Y \\
  5  & 130.425 & -0.831   & $(-36.4, -34.9)$  &  $(-31.7, -29.7)$  & C$|$C  &  Y \\
  ...
\enddata
\tablecomments{``Y'' = Yes, ``?'' = Possible, ``N'' = No. For the quality levels, ``A'' denotes the most reliable outflow candidates which are typically high-velocity ones, ``B'' denotes reliable outflow candidates (which include the majority of the outflow candidates), and ``C'' denotes outflow candidates whose characteristics were not obvious. Outflow candidates with scores of ``D'' were removed from our sample, owing to their ambiguous outflow characteristics. For more details of the quality levels see \citet{LLX2018}.\\
(This table is available in its entirety in machine-readable form. A portion is shown here for guidance.)}
\end{deluxetable}

\begin{deluxetable}{lccccccccc}
\tablecolumns{10}
\tabletypesize{\small}
\setlength\tabcolsep{2.8pt}
\tablecaption{Outflow candidate Sample Statistics \label{table:Statistics}}
\tablehead{
 \colhead{Sub-region} & \colhead{Numbers of Outflow} & \colhead{Numbers of} & \multicolumn{3}{c}{Quality Level} & \colhead{} & \multicolumn{3}{c}{WISE Detection}\\
 \cline{4-6} \cline{8-10}
 \colhead{} & \colhead{Total(Blue$|$Red)} & \colhead{Bipolar Outflow} & \colhead{A} & \colhead{B} & \colhead{C} & \colhead{} & \colhead{Y} & \colhead{?} & \colhead{N}
}
\startdata
Perseus arm      & 284(186$|$180)    & 82   &  41$|$14\%   & 160$|$56\%  & 83$|$29\%  & & 133$|$47\% & 136$|$48\% & 15$|$5\%  \\
Local arm        & 162(67$|$107)     & 12   &  10$|$6\%    & 79$|$49\%   & 73$|$45\%  & & 12$|$7\%   & 87$|$54\%  & 63$|$39\% \\
Interarm 1       & 5(2$|$3)          & 0    &  3$|$60\%    & 2$|$40\%    &  0$|$0\%   & & 2$|$40\%   & 3$|$50\%   & 0$|$0\%  \\
Interarm 2       & 3(2$|$3)          & 2    &  2$|$67\%    & 1$|$33\%    &  0$|$0\%   & & 1$|$33\%   & 2$|$67\%   & 0$|$0\%   \\
Outer arm        & 5(5$|$4)          & 4    &  2$|$40\%    & 2$|$40\%    &  1$|$20\%  & & 5$|$100\%  & 0$|$0\%    & 0$|$0\%   \\
Total            & 459(262$|$297)    & 100  & 58$|$13\%    & 244$|$53\%  & 157$|$34\% & & 153$|$33\% & 228$|$50\% & 78$|$17\% \\
\enddata
\tablecomments{See Table \ref{table:examples} for the description of quality level and WISE detection. In addition, we present the numbers and percentages and separate them with a ``$|$'' for quality level and WISE detection.}
\end{deluxetable}

Table \ref{table:Statistics} shows that the outflow candidates in the two interarm regions (interarms 1 and 2) contained only a minority of the total outflow candidates, indicating that star formation was primarily concentrated in the Galactic spiral arms. The quality level revealed that the interarm regions were excellent locations to study relatively isolated star formation. In addition, more bipolar outflow candidates were detected in the Perseus arm than in the Local arm. The outflow candidates in the Perseus arm also had a higher quality level and possessed a larger proportion of WISE associations than those in the Local arm. All these indicated that the Perseus arm contained more active star forming activities.

\subsection{Comparison with W5 Complex CO (3 $\rightarrow$ 2) Outflow Surveys}\label{subsection:comparison}

There are a number of outflow studies toward the W3/4/5 complex, such as \citep[e.g.,][]{BMR2002, GBW2011, ZZW2011}. Among these studies, \citet{GBW2011} conducted a CO (3 $\rightarrow$ 2) outflow survey toward the entire W5 complex (a total of $\sim$ 3 deg$^2$ with $136\degr.1 \lesssim l \lesssim 138\degr.7$ and $0\degr.3 \lesssim b \lesssim 2\degr.0$) with data acquired at the 15 m James Clerk Maxwell Telescope (JCMT) using the HARP array, and they detected 40 outflows. We therefore compared our work with their survey. With a final map HPBW resolution of 18$\arcsec$ and RMS of $\sim$ 0.06 -- 0.11 K at 0.42 km s$^{-1}$ channels, the sensitivity of the data used by \citet{GBW2011} was superior to that considered in the current study (HPBW is 49$\arcsec$, RMS is 0.45 K in a $0.16$ km s$^{-1}$ channel). After comparing with the outflow detection in L1448 from \citet{HFR2007}, \citet{GBW2011} concluded that they were able to detect any outflow, but might count fewer lobes and it was also difficult to make flow-counterflow associations. Their results (see their figure 7) also showed that they might miss candidate outflow lobes by a factor of $\sim$ 2.

\begin{figure}[!ht]
\centering
\includegraphics[height=0.9\textwidth,angle=-90]{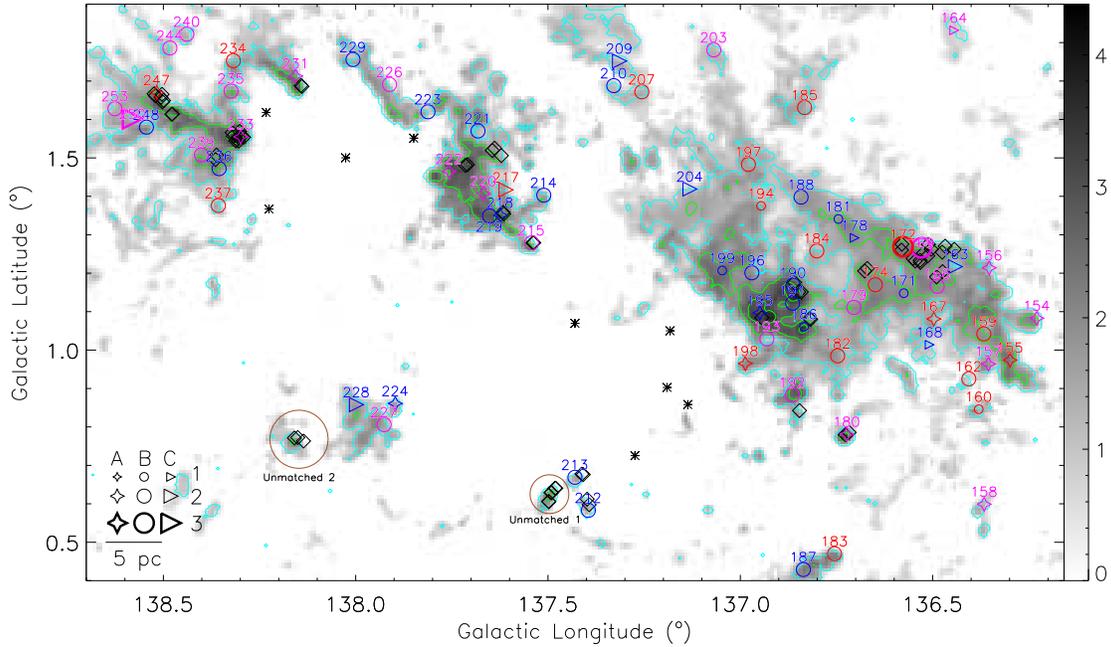}
\caption{Outflow candidate distribution in the W5 complex. The background gray-scale map is the integrated intensity map of \co in the range of [-62, -30] km s$^{-1}$, the gray value is the square root of the integrated intensity, and the color bar is in units of (K km s$^{-1}$)$^{1/2}$. The green contours are the integrated intensity map of $^{13}$CO in the same velocity range as the $^{12}$CO map. Their levels are 10, 30, 60, 90, 120 $\times$ 0.58 K km s$^{-1}$ (1$\sigma$). The cyan contours denote \xco emission boundaries, where the main beam brightness temperatures are larger than 3 $\times$ RMS in at least three successive channels. The blue/red open circles denote the blue/red lobes. The markers to describe quality level and classification of outflow candidates are placed in the bottom left corner of the panel, where 3, 2 and 1 denote high-, intermediate- and low-mass outflow candidates (see the criterion in Section \ref{subsection:parameters}), respectively. The magenta and red/blue colors of the shapes and the indexes denote bipolar outflow candidates and outflow candidates that have only a red/blue lobe, respectively. The diamonds represent the outflows found by \citet{GBW2011}. The brown circles denote the unmatched regions between (candidate) outflows in \citet{GBW2011} and this work. The stars indicate OB stars (see the description in Section \ref{subsection:distribution}). The physical scale bar is reported in the bottom left corner of the panel.}
\label{Fig:compare to W5}
\end{figure}

Figure \ref{Fig:compare to W5} superposes the outflow survey of \citet{GBW2011} to that from this work. Overall, the outflow candidates in this work shared a similar distribution to those in their survey, and 90\% of the outflows in their survey were detected. The circles marked by ``unmatched 1'' and ``unmatched 2'' denote the unmatched regions, where ``unmatched 1'' corresponds to outflows 18 and 19, and ``unmatched 2'' to outflows 39 and 40, in their survey. The sizes of the major axis of these four outflows were less than 23$\arcsec$, which could be the likely reason for our failure to detect them. In addition, CO (3 $\rightarrow$ 2) lines require warmer temperatures for excitation ($E_{\nu}/k = 33.2$ K above ground states) and may show higher detection rates in warmer regions than CO (1 $\rightarrow$ 0) lines \citep[e.g.,][]{TSO2008, CRS2010}. One can consult figures 3 and 4 in \citet{NKK2011} as examples that demonstrate the difference between CO (1 $\rightarrow$ 0) and CO (3 $\rightarrow$ 2) outflows.

\begin{figure}[!ht]
\centering
\includegraphics[width=0.246\textwidth,angle=-90]{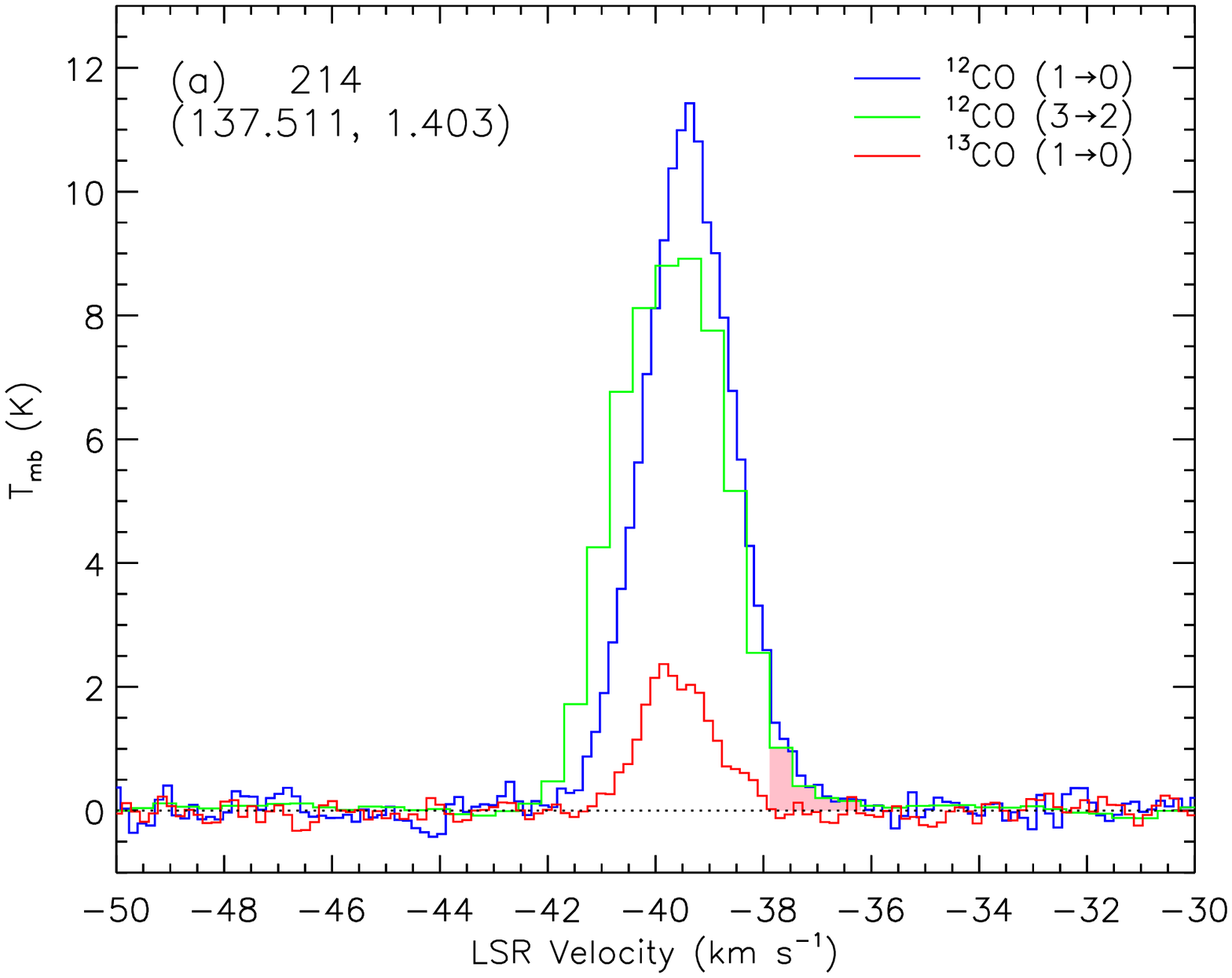}
\includegraphics[width=0.246\textwidth,angle=-90]{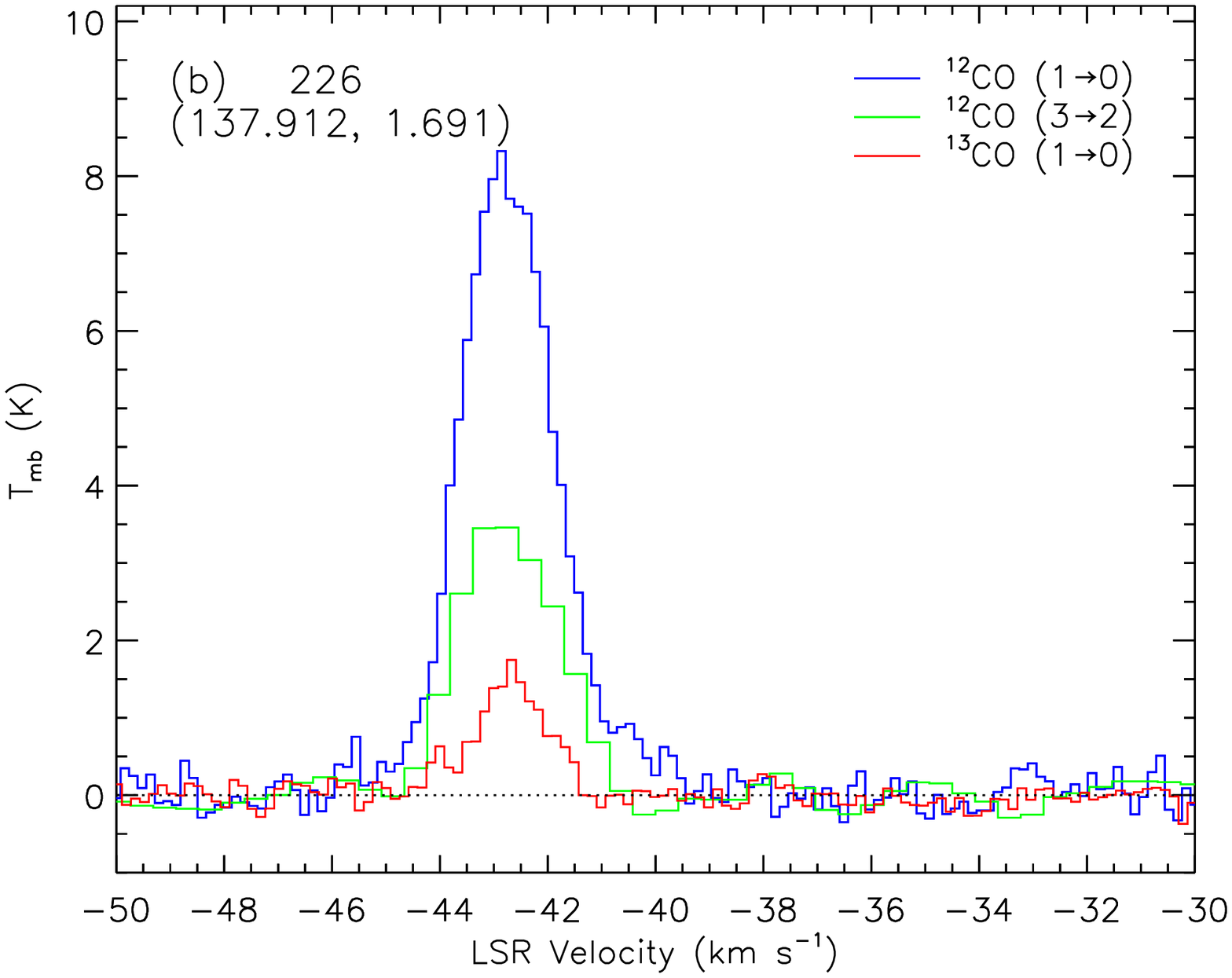}
\includegraphics[width=0.246\textwidth,angle=-90]{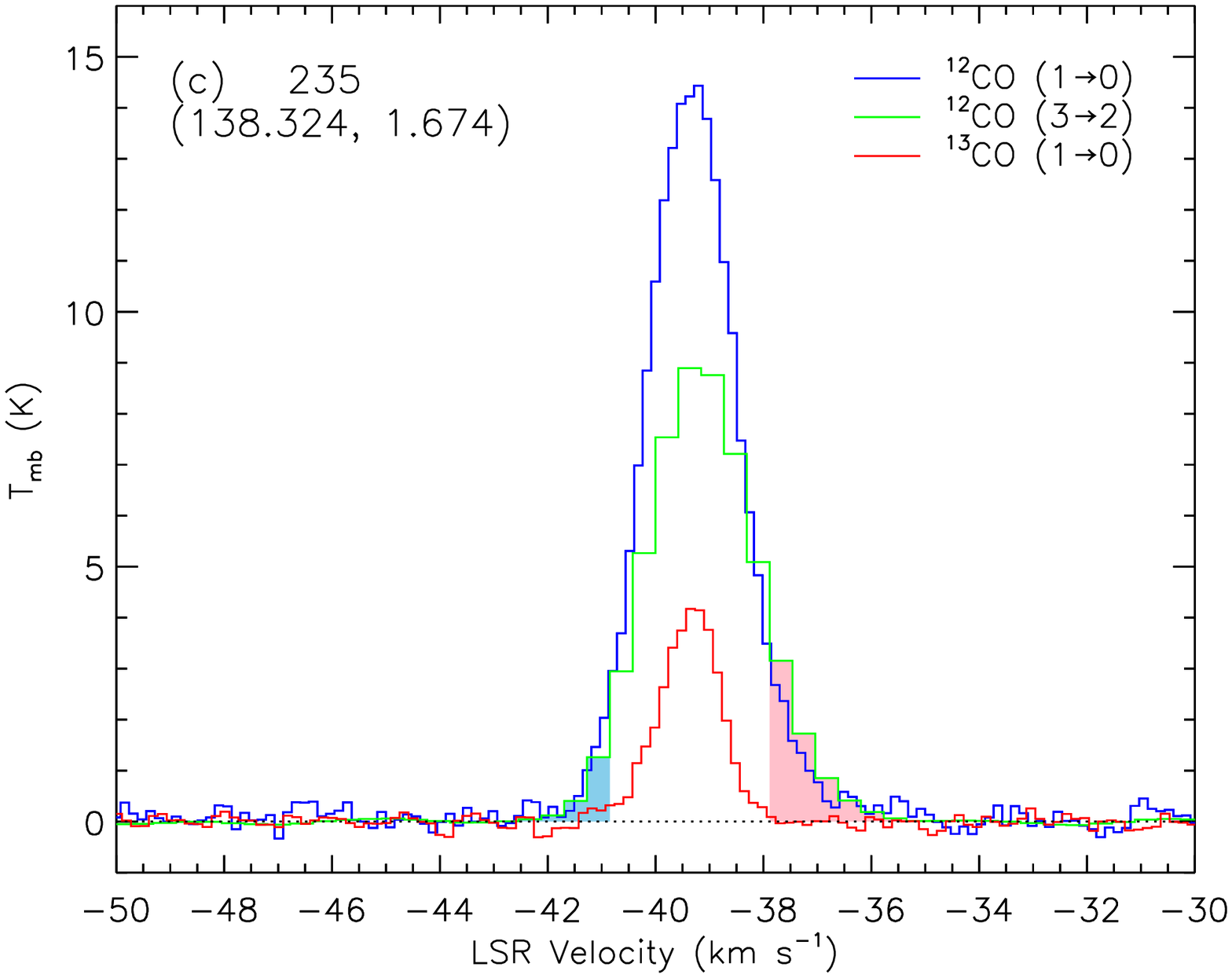}
\includegraphics[width=0.35\textwidth,angle=-90]{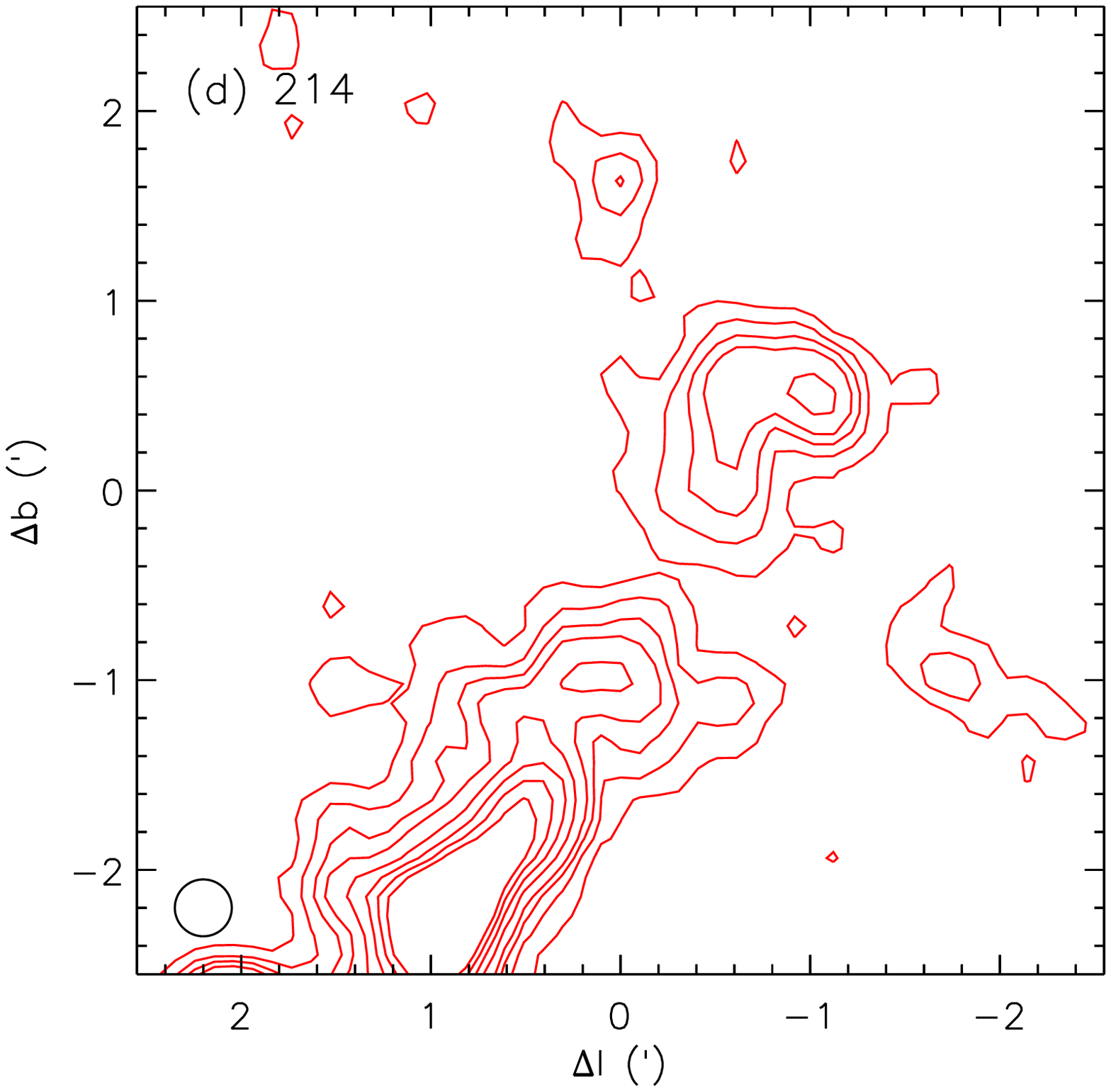}
\includegraphics[width=0.35\textwidth,angle=-90]{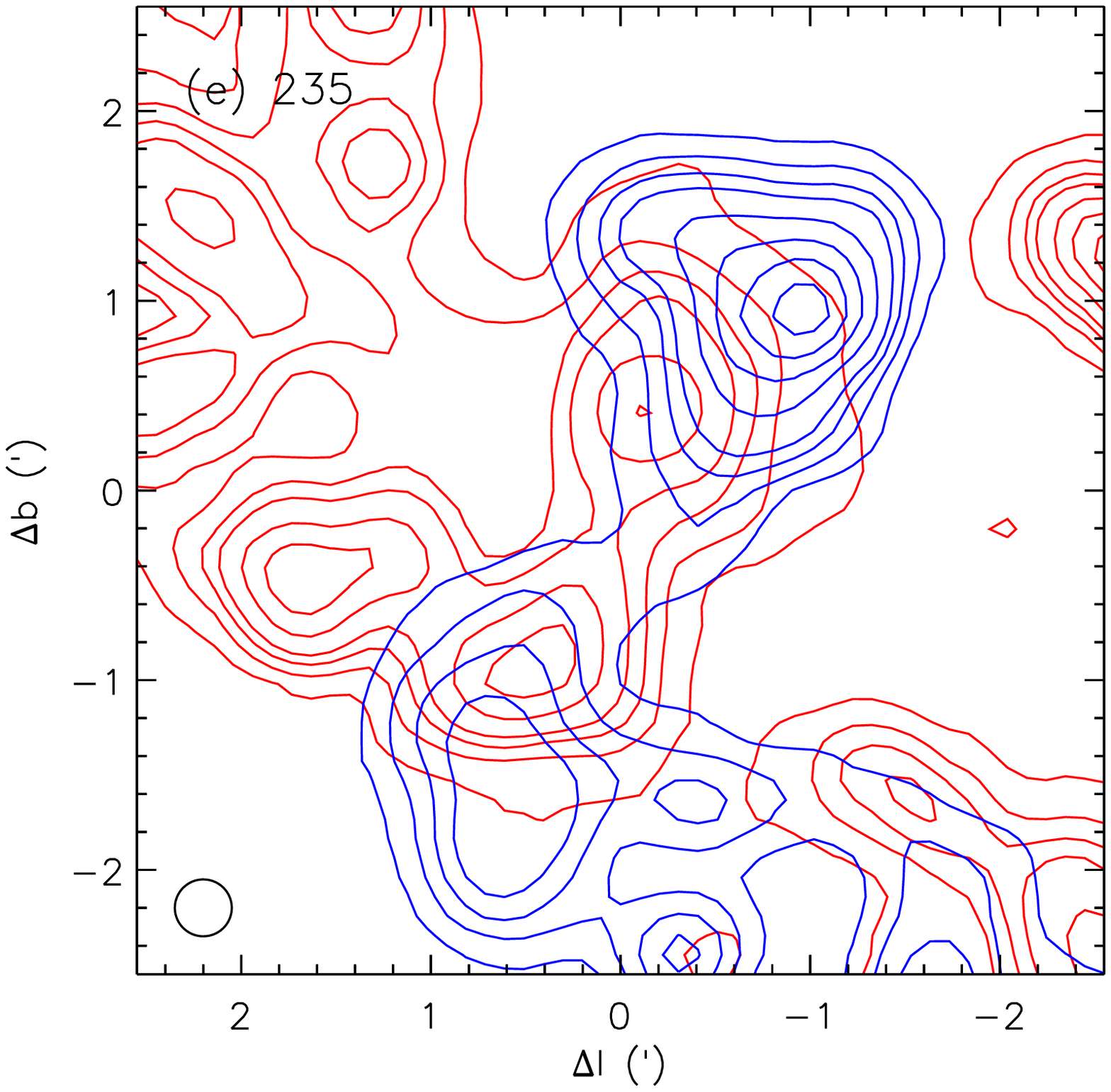}
\caption{(a) -- (c): mean spectra of \co (1 $\rightarrow$ 0), \co (3 $\rightarrow$ 2) and \xco (1 $\rightarrow$ 0) averaged over $3 \times 3$ pixels centered at the position shown in the spectra using Galactic coordinates. The possible velocity range of each high-velocity line wing of \co (3 $\rightarrow$ 2) is specified with shading. Indexes of outflow candidates in this work are reported in the top left corner of panels (a) -- (e). (d) and (e): blue/red integrated intensity contours (integrated over the velocities indicated in upper panels with the same indexes) of the possible blue/red high-velocity line wings of \co (3 $\rightarrow$ 2). The contour levels are 0.5, 1, 1.5, 2, 3, 4, 5, 6 K km s$^{-1}$. The coordinate origin of panels (d) and (e) refers to the positions shown in panels (a) and (c), respectively. The circle in the lower left corner in panels (d) or (e) shows HPBW (18$\arcsec$) of \co (3 $\rightarrow$ 2) whose diameter is corresponding to $\sim0.17$ pc with distance of $\sim$ 1960 pc.}
\label{Fig:spectra}
\end{figure}

There were many outflow candidates that were not reported in \citet{GBW2011}. One reason may be the difference between the CO (3 $\rightarrow$ 2) and (1 $\rightarrow$ 0) lines (see the panels (a) and (b) of Figure \ref{Fig:spectra}). The CO (3 $\rightarrow$ 2) presented a possible red high-velocity line wing while the CO (1 $\rightarrow$ 0) showed a blue one (e.g., Figure \ref{Fig:samples} and the panels (a) and (d) in Figure \ref{Fig:spectra}). Another reason may be that they were likely to miss many outflow candidates with large size (the mapped size in their studies was $\sim$ 2$\arcmin$) and/or in complex environments (e.g., the panel (d) in Figure \ref{Fig:spectra}). The CO (3 $\rightarrow$ 2) lines were likely to present more complex structures (e.g., Figure \ref{Fig:samples} and the panel (e) in Figure \ref{Fig:spectra}) or absent some structures \citep[see an example close to the regions labeled by ``EL 32'' and ``LFAM 36'' in figures 3 and 4 in][]{NKK2011} relative to CO (1 $\rightarrow$ 0) lines.

\subsection{Calculation of the Outflow Candidates' Parameters} \label{subsection:parameters}

The area of an outflow lobe candidate, $A_\mathrm{lobe}$, is typically estimated by the contour that contains 40\% of its peak integrated intensity. However, if an outflow candidate is contaminated by other components but can be resolved, $A_\mathrm{lobe}$ is estimated by the maximum area of the vicinity of the lobe where the outflow candidate's components can be distinguished from the cloud. Owing to the limited capacity to resolve the detailed structures of an outflow candidate, the length of an outflow lobe candidate, $l_{\mathrm{lobe}}$, is estimated with an average collimation factor (2.45) derived by \citet{WWZ2004} from 213 outflows. The collimation factor is defined as the ratio of the major and minor diameters (denoted them as $a$ and $b$, respectively, $a = 2.45 b$) of the region of an outflow lobe candidate. Therefore, $A_\mathrm{lobe} \sim \pi ab/4$ and $l_{\mathrm{lobe}} \sim a \sim 1.77 \sqrt{A_\mathrm{lobe}}$.

The column density is estimated with \co assuming local thermal equilibrium (LTE), an area filling factor $f = 1$, and optically thin line wings using $N(^{12}\mathrm{CO})=1.5\times 10^{15}\int{T_{\mathrm{mb}}}dv$ with $T_{ex} = 30$ K \citep[see][]{SSS1984}. If the \xco emission is above the noise (i.e., $>$ RMS) over the line wing velocity range of the outflow lobe candidate, the correction factor for the opacity is $f_{\tau} = \tau_{12}/(1-e^{\tau_{12}})$ and $N(^{12}\mathrm{CO})$ is corrected to be $f_{\tau} N(^{12}\mathrm{CO})$, where $\tau_{12}$ is the optical depth of the \co line. In such a situation, $\tau_{12}$ can be calculated by multiplying $\tau_{13}$ (the optical depth of the \xco line) by the abundance ratio, $[^{12}\mathrm{CO}]/[^{13}\mathrm{CO}] \sim 70$ \citep{MSB2005} assuming that \co is optical thick, \xco is optical thin, LTE, and identical filling factors (i.e., 1) and excitation temperatures (i.e., 30 K) for both isotopologues. The factor $\tau_{13}$ therein equals $-\ln(1-T^{13}_\mathrm{mb}/26.6)$ \citep[for an excitation temperature of 30 K,][]{KOY1998}, where $T^{13}_\mathrm{mb}$ is the main beam brightness temperature of \xco line. The total column density of an outflow lobe candidate of H$_2$ gas, $N_{\mathrm{lobe}}$, can be calculated by multiplying the corrected $N(^{12}\mathrm{CO})$ by a conversion factor $[\mathrm{H}]/[^{12}\mathrm{CO}]$ $=1\times 10^4$ \citep[see e.g., ][]{SSS1984}.

The mass of the outflow lobe candidate reads $M_{\mathrm{lobe}}=(N_{\mathrm{lobe}}\times A_{\mathrm{lobe}})m_{\mathrm{H_2}}$, where $m_{\mathrm{H_2}}$ is the mass of a hydrogen molecule. The momentum, $P_{\mathrm{lobe}}$, and kinetic energy, $E_{\mathrm{lobe}}$, of an outflow lobe candidate are based on the outflow candidate's velocity relative to the central cloud (the lobe velocity, $\langle\Delta v_{\mathrm{lobe}}\rangle$). These physical parameters are
\begin{gather*}
 \langle\Delta v_{\mathrm{lobe}}\rangle=\frac{\sum_{i}\left(v_i-v_{\mathrm{peak}}\right)T_i\Delta v_{\mathrm{res}}}{\sum_{i}T_i\Delta v_{\mathrm{res}}}, \\
 \langle\Delta v^2_{\mathrm{lobe}}\rangle=\frac{\sum_{i}\left(v_i-v_{\mathrm{peak}}\right)^2T_i\Delta v_{\mathrm{res}}}{\sum_{i}T_i\Delta v_{\mathrm{res}}},\\
 P_{\mathrm{lobe}}=\displaystyle\sum_{A_{\mathrm{lobe}}}M_{\mathrm{lobe}}\langle\Delta v_{\mathrm{lobe}}\rangle,\\
 E_{\mathrm{lobe}}=\frac{1}{2}\displaystyle\sum_{A_{\mathrm{lobe}}}M_{\mathrm{lobe}}\langle\Delta v^2_{\mathrm{lobe}}\rangle,
\end{gather*}
where $i$ and $T_i$ are the channel index of the blue/red wing and the main beam brightness temperature of each channel, respectively, and $\Delta v_{\mathrm{res}}$ is the velocity resolution of the channel.

The total mass, momentum and energy of the outflow lobe candidates in the W5 complex (see Figure \ref{Fig:compare to W5}) were $M_\mathrm{tot} \sim 72$ $M_{\sun}$, $P_\mathrm{tot} \sim 193$ $M_{\sun}$ km s$^{-1}$, and $E_\mathrm{tot} \sim 6 \times 10^{45}$ erg. These values were much higher than those reported in \citet{GBW2011} whose corresponding values were $\sim 1.5$ $M_{\sun}$, $\sim 10.9$ $M_{\sun}$ km s$^{-1}$, and $\sim 7 \times 10^{44}$ erg, respectively. Considering only A-rated outflow candidates, the values were $\sim 18$ $M_{\sun}$, $\sim 61$ $M_{\sun}$ km s$^{-1}$, and $\sim 2 \times 10^{45}$ erg, respectively. The mass, momentum and energy ratios between the A-rated outflow candidates in this work and those in their survey were $\sim$ 12, $\sim$ 6 and $\sim$ 3. These were similar to the mean ratios for the (candidate) outflows that both detected in this work and in their survey, where the selected relatively isolated outflow candidates were 180, 212, 213, 215 and 231 (corresponding to outflows 10, 16, 17, 20 and 25 in their survey, respectively). The specific value decreased from mass ratio to momentum ratio and from momentum ratio to energy ratio, likely implying that this work exhibited a bias towards detecting mass at lower velocities. Therefore the mass and momentum in this work was higher than those reported in \citet{GBW2011} because lower-velocity gas was likely to dominate the column density in outflow lobes. Another reason for obtaining a higher mass and momentum in this work was possible that the outflows' parameters reported in their work did not include optical depth correction for the CO (3 $\rightarrow$ 2) lines \citep[the correction factor ranged from 1.8 to 14.3,][]{CRS2010}.

The dynamical timescale of an outflow lobe candidate is measured by $t_\mathrm{lobe} = l_\mathrm{lobe}/ \Delta v_\mathrm{max}$, where $\Delta v_\mathrm{max}$ is the maximum velocity of an outflow lobe candidate \citep[similar to][]{BSS2002} which is based on the sensitivity of our data, i.e., 0.45 K per 61 kHz channel for $^{12}$CO. The luminosity of an outflow lobe candidate is then $L_{\mathrm{lobe}}=E_{\mathrm{lobe}}/t_{\mathrm{lobe}}$. The physical properties of the outflow candidates in the Perseus arm, the Local arm and interarm 1 are listed in Table \ref{Table:physical properties}. Outflow candidates in interarm 2 and the Outer arm are not included in the table because they are highly clustered\footnote{The mean angular outflow size ($\lesssim 21\arcsec$ at $\gtrsim 4$ kpc) expected from outflow surveys in \citet{ABG2010} and \citet{LLQ2015} is far less than the HPBW of the current study \citep[$\sim 49\arcsec$, see][]{LLX2018}.} and hence causes large errors in $A_\mathrm{lobe}$ and other related physical parameters \citep[see more details in][]{LLX2018}.

\begin{deluxetable}{llcccccccccc}
\centering
\setlength\tabcolsep{3pt}
\tablecolumns{12}
\tabletypesize{\small}
\tablewidth{0pt}
\tablecaption{Physical Properties of the Outflow Candidate Samples\label{Table:physical properties}}
\tablehead{
\colhead{Index} & \colhead{Lobe} & \colhead{$l$} & \colhead{$b$} & \colhead{$v_\mathrm{c}$} & \colhead{$\langle\Delta v_{\mathrm{lobe}}\rangle$} & \colhead{$l_{\mathrm{lobe}}$} & \colhead{$M_{\mathrm{lobe}}$} & \colhead{$P_{\mathrm{lobe}}$} & \colhead{$E_{\mathrm{lobe}}$} & \colhead{$t_\mathrm{lobe}$} & \colhead{$L_{\mathrm{lobe}}$} \\
 \colhead{} & \colhead{} & \colhead{($\degr$)} &  \colhead{($\degr$)} & \colhead{(km s$^{-1}$)} & \colhead{(km s$^{-1}$)} & \colhead{(pc)} & \colhead{($M_{\odot}$)} & \colhead{($M_{\odot}$ km s$^{-1}$)} & \colhead{(10$^{43}$ erg)} & \colhead{(10$^5$ yr)} & \colhead{(10$^{30}$ erg s$^{-1}$)}
}
\startdata
\multicolumn{12}{l}{the Perseus arm}\\
\hline
  1 & Blue & 130.136 & -4.467 & -41.68 &  2.10  & 1.56 &  0.23    & 0.49   &  1.02    &  5.82    & 0.55     \\
  2 & Blue & 130.155 & -4.537 & -42.65 &  1.43  & 1.98 &  0.36    & 0.51   &  0.72    &  9.94    & 0.23     \\
  3 &  Red & 130.399 & -0.728 & -31.84 &  2.60  & 1.87 &  0.60    & 1.55   &  4.01    &  5.30    & 2.40     \\
  4 &  Red & 130.401 &  1.667 & -42.86 &  1.80  & 0.95 &  0.05    & 0.09   &  0.16    &  3.79    & 0.14     \\
  5 & Blue & 130.417 & -0.842 & -33.79 &  2.01  & 1.53 &  0.31    & 0.62   &  1.23    &  5.74    & 0.68     \\
    &  Red & 130.433 & -0.819 & -33.34 &  2.44  & 5.65 &  9.22    & 22.50  &  54.20   &  15.20   & 11.30    \\
  ...
\enddata
\tablecomments{$v_\mathrm{c}$ denotes the  center velocity of \xco for an outflow lobe candidate.\\
(This table is available in its entirety in machine-readable form. A portion is shown here for guidance.)}
\end{deluxetable}

In addition to the opacity, inclination and blending can also affect the derived outflow parameters \citep{AG2001B}. Inclination effects can greatly reduce the amount of $l_{\mathrm{lobe}}$, $\langle\Delta v_{\mathrm{lobe}}\rangle$, $P_{\mathrm{lobe}}$, $E_{\mathrm{lobe}}$, $t_{\mathrm{lobe}}$ and $L_{\mathrm{lobe}}$. Assuming a random distribution of inclination angles, the mean value is given by $\langle\theta\rangle=\int^{\pi/2}_{0}\theta \sin \theta d\theta=57\degr.3$ \citep{BATC1996, DAM2014}. The blending effect can reduce the amount of $M_{\mathrm{lobe}}$ and other physical parameters related to $M_{\mathrm{lobe}}$ (e.g., $P_{\mathrm{lobe}}$, $E_{\mathrm{lobe}}$, and $L_{\mathrm{lobe}}$) by a factor of $\sim 2.0$, as previous studies have shown \citep{ML1985, ABG2010, NSB2012}. The effect that strong shocks dissociate molecular gas into atomic material only introduces a small error to the calculated mass and its related physical parameters, and it is equally true for the assumption of a single constant temperature \citep[e.g.,][]{DR1999, DC2007, CRS2010}. For more information regarding the correction factors for the inclination and blending see the summary in table 5 of \citet{LLX2018}. Caution was taken that those corrections were only applied to statistics. We do not consider inclination or blending effects in this work except where otherwise noted.

The outflow candidates were classified as high-, intermediate- and low-mass for the Perseus arm, the Local arm and interarm 1 using the mass criteria of \citet{YTU2018}. An outflow candidate was classified as high-mass if the corrected mass for any one lobe was greater than 5 $M_{\sun}$. An intermediate-mass classification was given to candidates whose corrected mass ranged from 0.5 -- 5 $M_{\sun}$, while a low-mass designation was assigned to the rest.

\subsection{The Limitations of the Survey}\label{Section:limit}

Because of the sensitivity of 0.45 K per 61 kHz channel (corresponding to $\sim$ 0.16 km s$^{-1}$ at 115.271 GHz) and moderate HPBW of 49$\arcsec$ for $^{12}$CO, some outflows with faint emissions, low velocities, complex environments or small sizes could be missed. Some clustered outflows might be presented as extended lobes because of poor resolution and might be omitted. The identified outflow candidates might actually be multi-outflows owing to the moderate resolution, or highly clustered in regions with distance $\gtrsim$ 3.8 kpc(see Section \ref{subsection:parameters}). Some outflow candidates were regarded as monopolar because their counterflows could be too faint/confused to be detected.

The total mass, momentum and energy of the outflow lobe candidates in the W5 complex (see Section \ref{subsection:parameters}) are $\sim$ 2 -- 3 times higher than those in the Perseus molecular cloud complex \citep[their corresponding parameters were $\sim$ 26 $M_{\sun}$, $\sim$ 60 $M_{\sun}$ km s$^{-1}$, and $\sim$ 2 $\times$ $10^{45}$ erg, see][]{ABG2010}. The (candidate) outflow mass ratio between those in W5 and those in the Perseus molecular cloud complex (i.e., $\sim$ 2 -- 3) was less than their corresponding gas mass ratio \citep[i.e., $\sim$ 3 -- 5,][]{BWJ2008, GBW2011}. Two factors might be responsible for the deficiency of the mass, momentum and energy of the outflow candidates: a greater fraction of the outflow mass was blended with the cloud in the more turbulent W5 region \citep{GBW2011}, and/or the limiting mass, momentum and energy of the outflow candidates in the Perseus arm were higher than those in the Perseus molecular cloud complex.

\begin{figure}[!ht]
\centering
\vspace{0.2cm}
\hspace{0.08cm}
\includegraphics[width=0.27\textwidth]{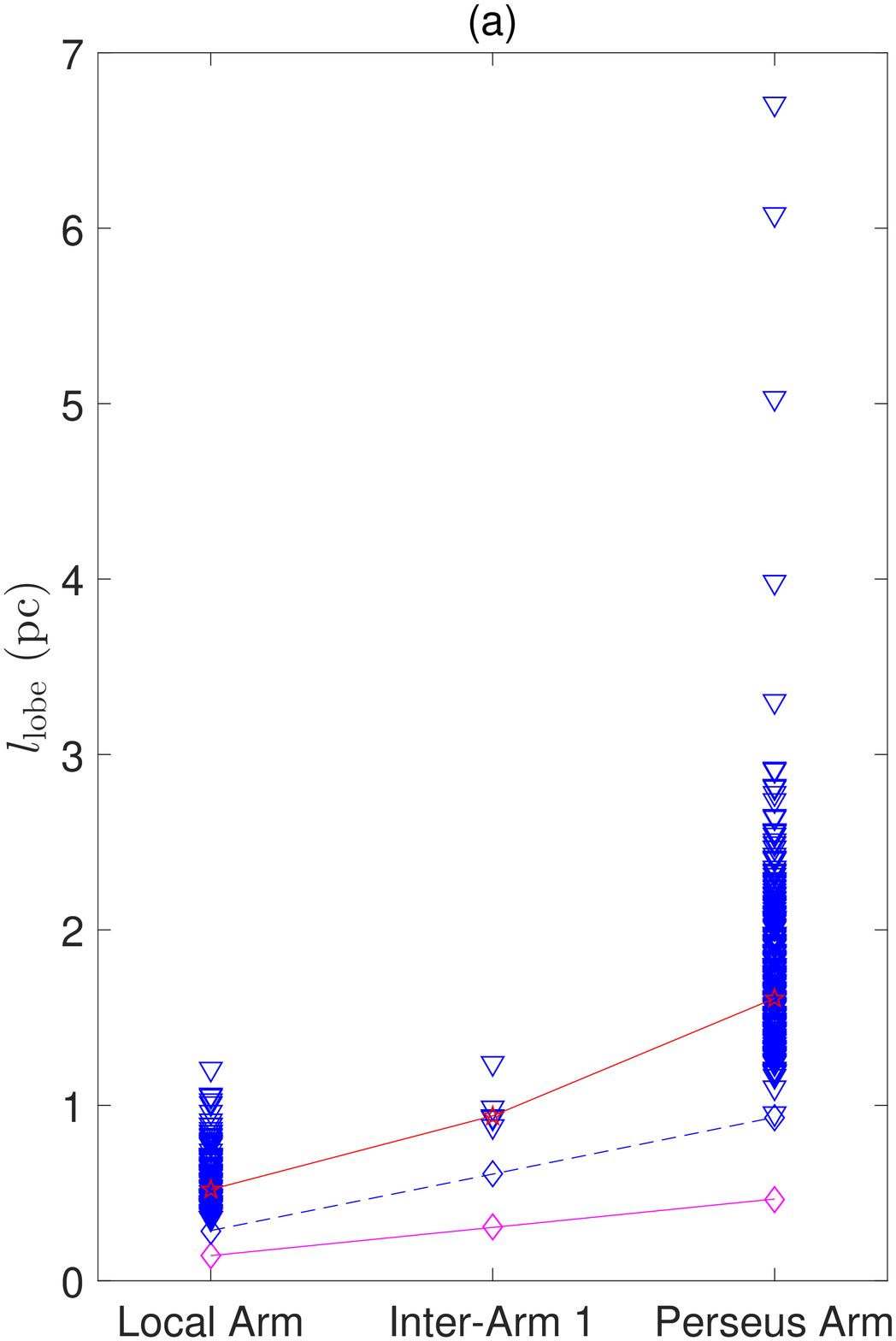} \;
\includegraphics[width=0.28\textwidth]{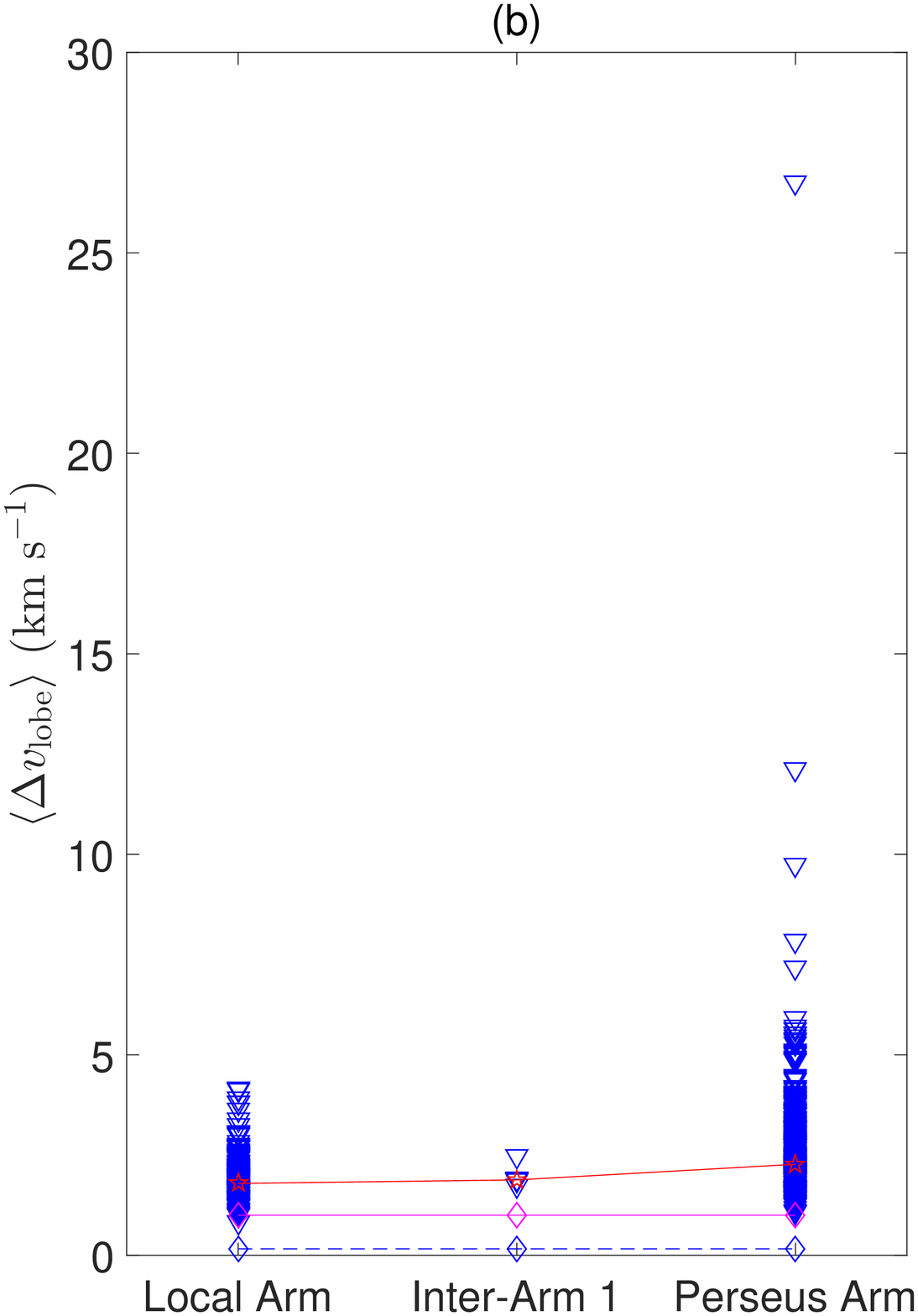}\;
\includegraphics[width=0.28\textwidth]{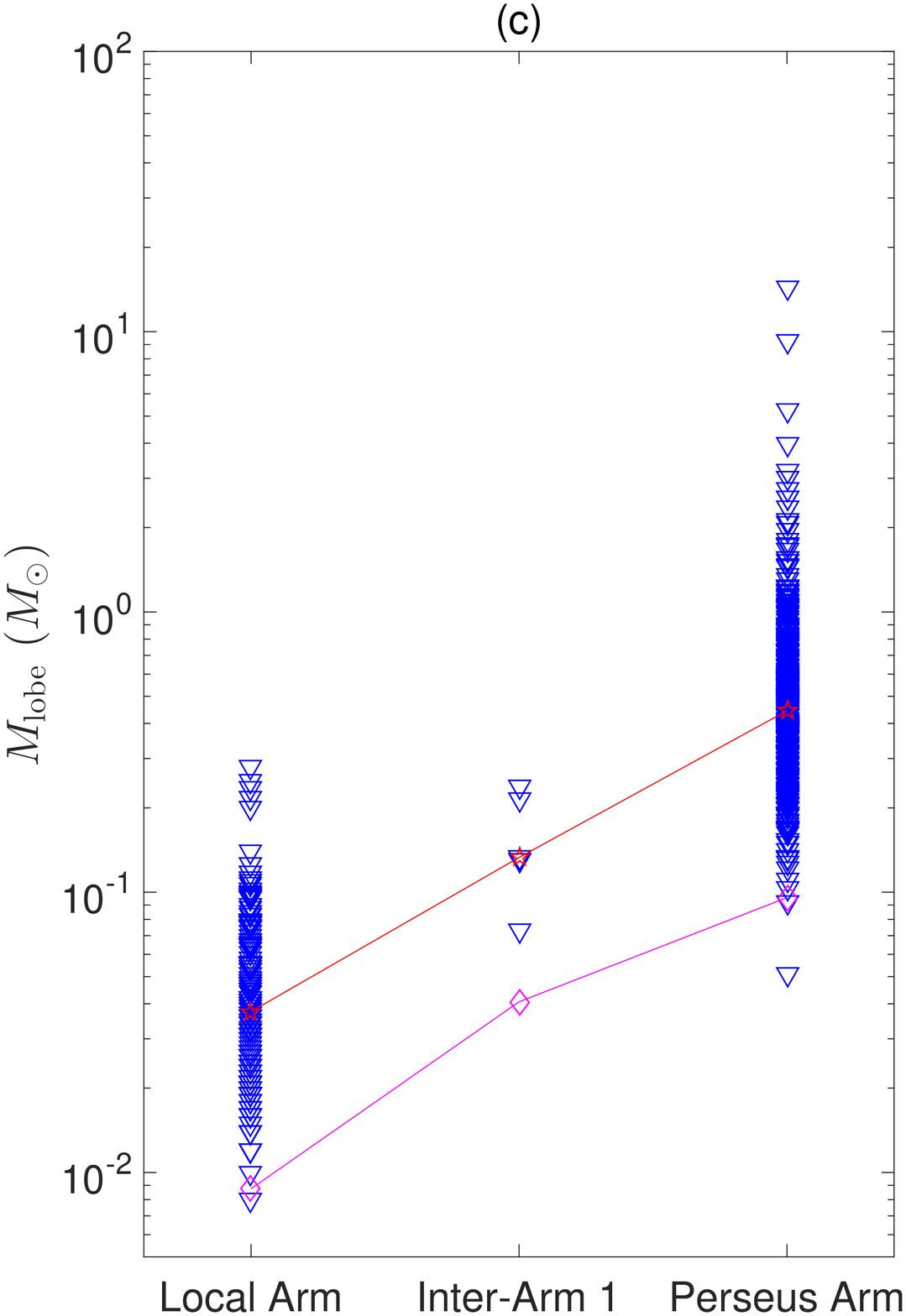} \\
\vspace{0.5cm}
\includegraphics[width=0.28\textwidth]{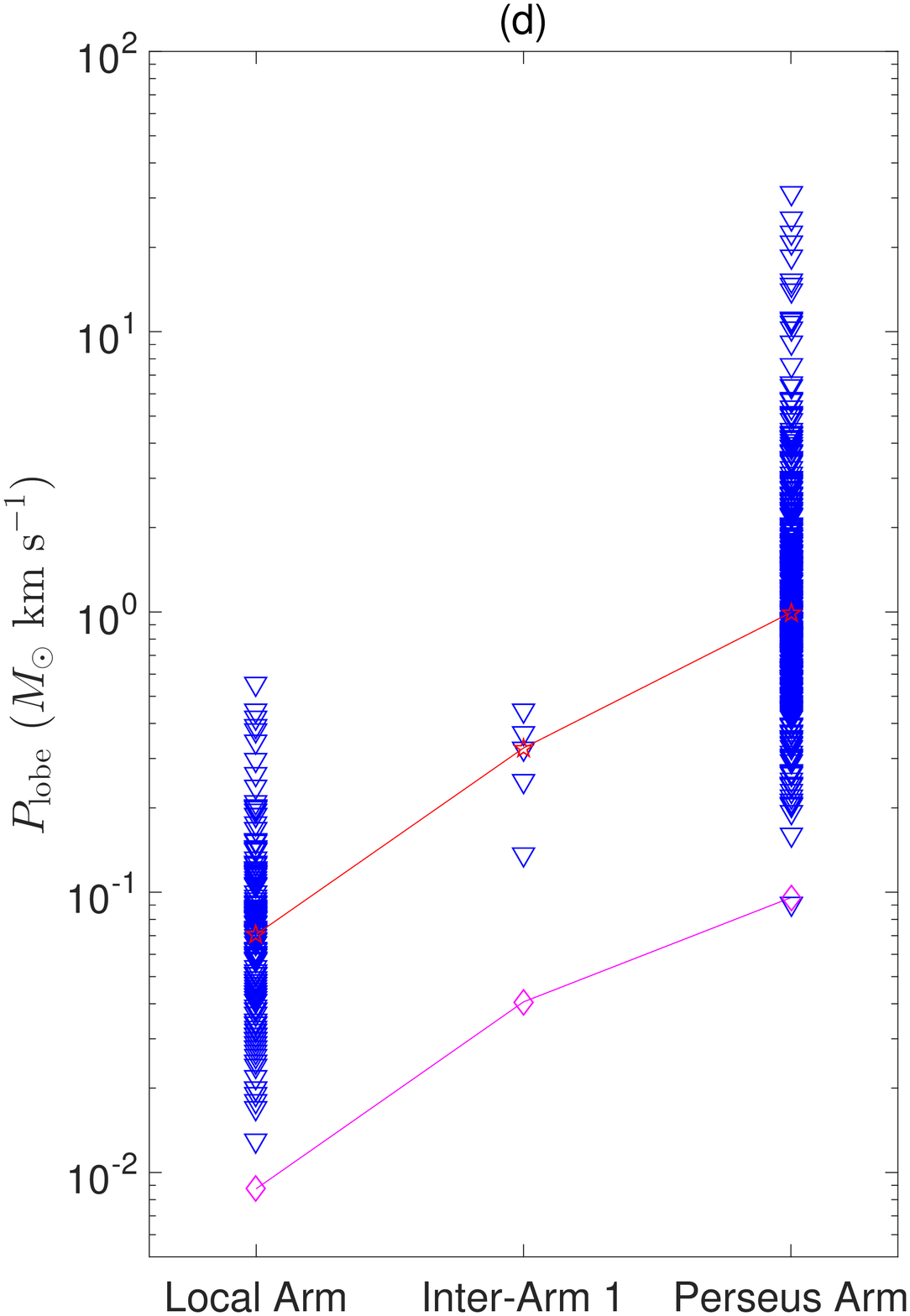}\;
\includegraphics[width=0.28\textwidth]{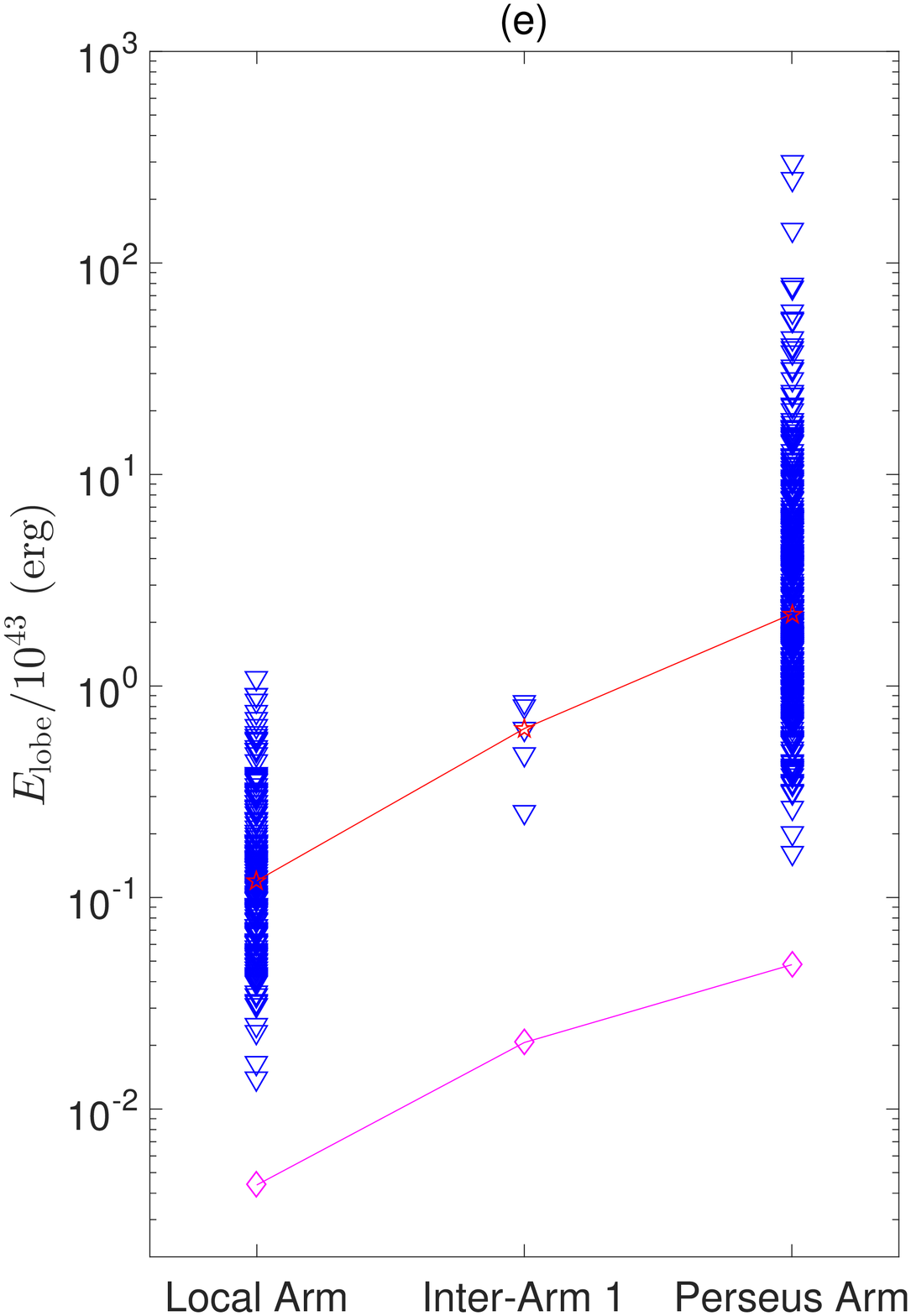}\;
\includegraphics[width=0.28\textwidth]{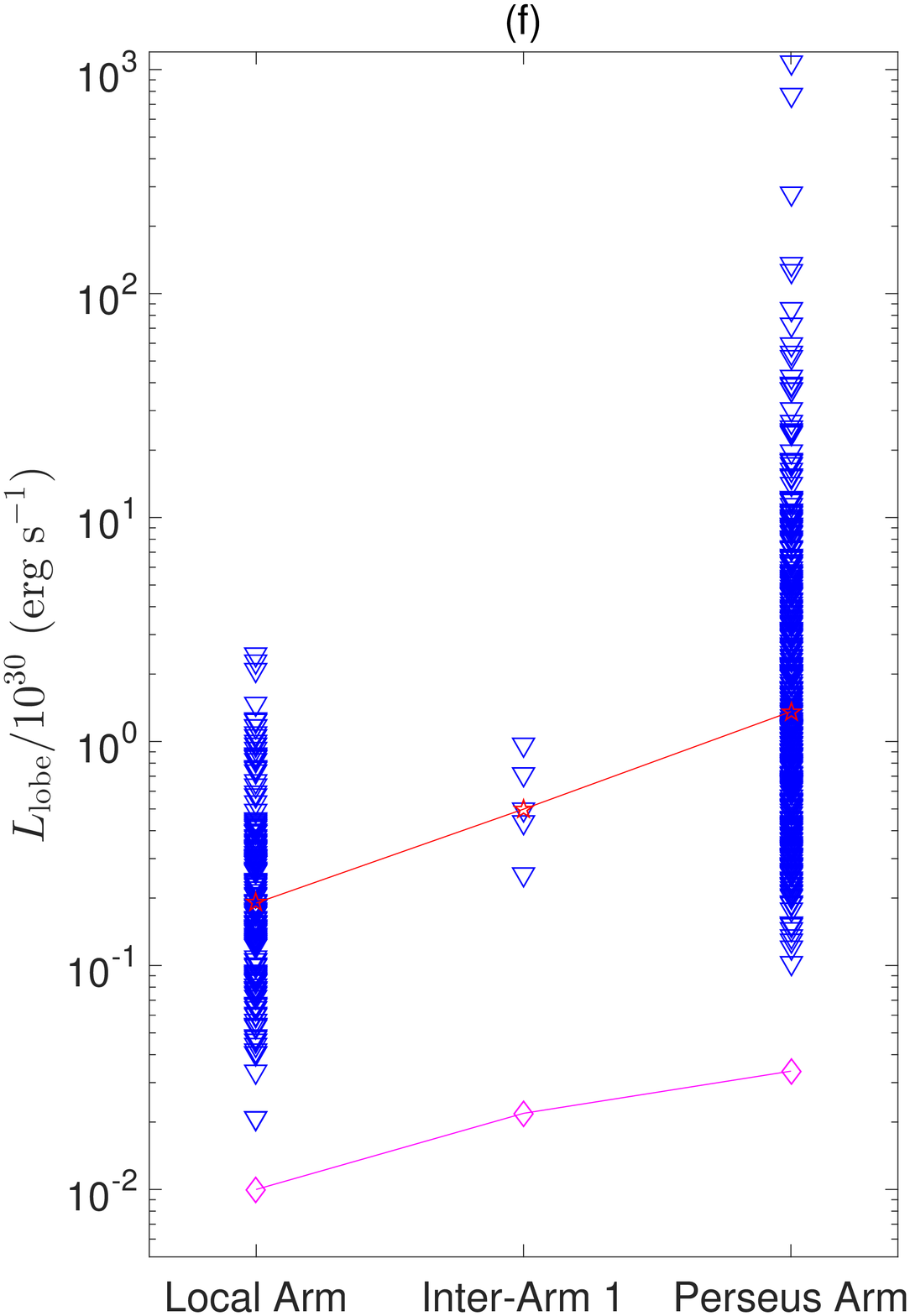}
\caption{Physical parameters of the candidate outflow lobes in Perseus arm, the Local arm and the interarm 1. The red lines and open pentagrams show the median values for all panels. (a) The lengths of the candidate outflow lobes, where the purple and blue line/diamonds denote the HPBW of $^{12}$CO and three pixel sizes, respectively. (b) The velocities of the candidate outflow lobes, where the dash line shows the velocity resolution (0.16 km s$^{-1}$) for $^{12}$CO, and the purple line indicates 1.0 km s$^{-1}$, which refers to the threshold of the extensional velocity for a minimum velocity extent relative to the ambient gas \citep[see][]{LLX2018}. (c)--(f): The mass, momentum, energy and luminosity, respectively, of the candidate outflow lobes. The purple lines and diamonds denote the sensitivities of this work, where the HPBW, the velocity of candidate outflow lobe (1.0 km s$^{-1}$) and a main beam brightness temperature of 3 $\times$ RMS were used to calculate the sensitivities.}
\label{Fig:statistic}
\end{figure}

Figure \ref{Fig:statistic} shows the physical parameters of the candidates in the Perseus arm, the Local arm and interarm 1. The lobe candidate velocity, $\langle\Delta v_{\mathrm{lobe}}\rangle$, was less than the extensional velocity of a minimum velocity extent relative to the ambient gas. We saw a $\langle\Delta v_{\mathrm{lobe}}\rangle$ that was lower than 1.0 km s$^{-1}$ for some sample points in panel (b). In panel (c), because the main beam brightness temperatures in some channels were less than 3 $\times$ RMS over the line wing velocity range of the outflow lobe candidates, we saw smaller candidate lobe mass ($M_\mathrm{lobe}$) than the limiting mass for some sample points. $\langle\Delta v_{\mathrm{lobe}}\rangle$ was less affected by distance effect.

From Figure \ref{Fig:statistic}, the length, mass, momentum, energy and luminosity of the majority of the outflow lobe candidates in the Perseus arm were much greater than those in the other two regions (the Local arm and interarm 1). This implied that the amount of outflow missed in the Perseus arm was larger than those in the other two regions (i.e., we were more likely to see outflow clusters as single candidates).

Now we will discuss how many outflows we may have missed. From Figure \ref{Fig:compare to W5}, there are seven CO (3 $\rightarrow$ 2) outflows \citep[26 -- 32, from a total of 8 blue/red lobes,][]{GBW2011} near the densest region of the W5 complex, the bipolar outflow candidate 233. Combining the fact that \citet{GBW2011} might have missed candidate outflow lobes by a factor of $\sim 2$ (see Section \ref{subsection:comparison}), we may have missed possible outflow lobes by a factor \citep[scale-up factor, SUF, see ][]{LLX2018} of $\lesssim$ 8 in the Perseus arm. This value is similar to the SUFs used for the Gem OB1 molecular cloud complex, where a different method was applied by comparing the outflow column densities $N_\mathrm{o}$ \citep{LLX2018}. It is expected that the SUFs for the Local arm and interarm 1 were also $\lesssim$ 8 (see above). In addition, because $N_\mathrm{o}$ in the Local arm was $\sim$ 6 times higher than that in the Perseus arm, the SUF in the Local arm was probably less than $\sim$ 2.

\section{The Spatial Distribution of The Outflow Candidates}

In this section the spatial distribution of the outflow candidates in each arm or inter-arm region are described. Further steps were subsequently made to present close-up views of interesting active regions (IARs hereafter) in the W3/4/5 complex. The maps are shown in Figures \ref{Fig:compare to W5} and \ref{Fig:the Perseus arm} -- \ref{Fig:outer}.

\subsection{The Perseus Arm --- the W3/4/5 Complex} \label{subsection:distribution}

\begin{figure}[!ht]
\centering
\includegraphics[width=0.77\textwidth]{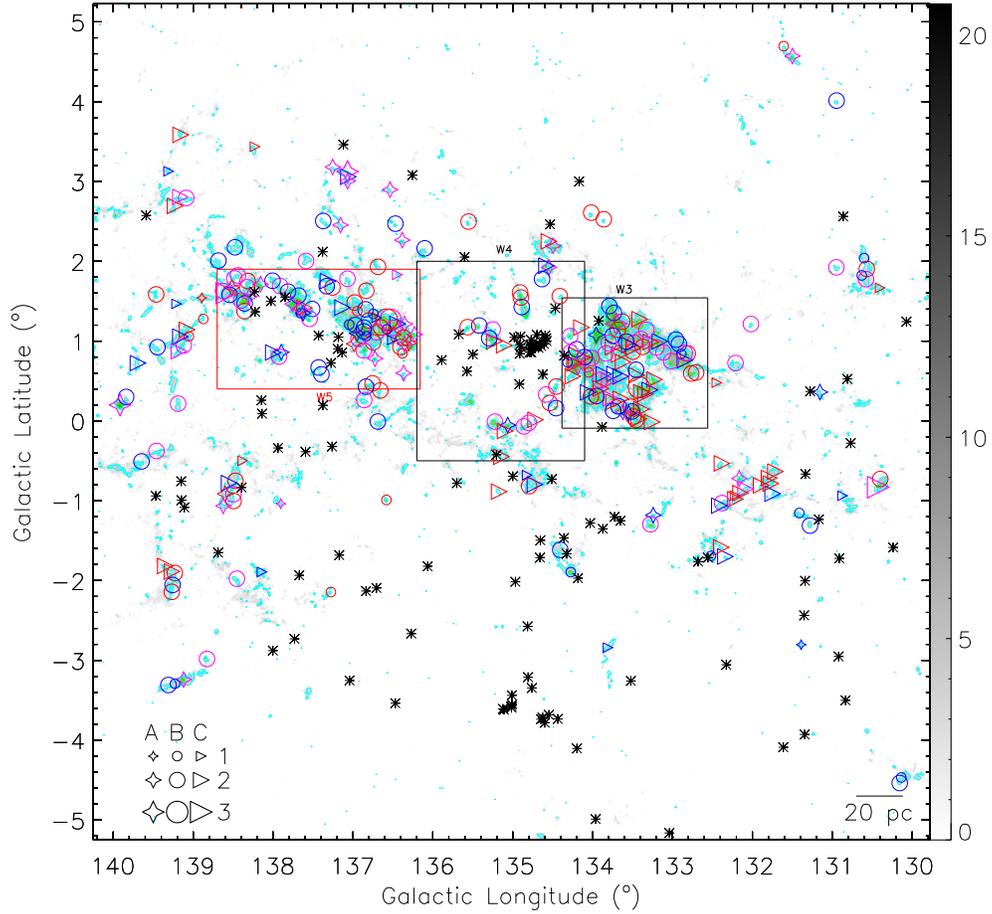}
\caption{Outflow candidate distribution in the Perseus arm. The background gray-scale map is the integrated intensity map of \co in the range of [-62, -30] km s$^{-1}$, the gray value is the square root of the integrated intensity, and the color bar is in units of (K km s$^{-1}$)$^{1/2}$. The green contours are the integrated intensity map of $^{13}$CO in the same velocity range as the $^{12}$CO map. Their levels are 10, 30, 60, 90 120 $\times$ 0.58 K km s$^{-1}$ (1$\sigma$). The cyan contours denote \xco emission boundaries, where the main beam brightness temperatures are larger than 3 $\times$ RMS in at least three successive channels. The blue/red open circles denote the blue/red lobes. The markers to describe quality level and classification of outflow candidates is placed in the bottom left corner of the panel, where 3, 2 and 1 denote high-, intermediate- and low-mass outflow candidates, respectively. The magenta, red/blue colors of the shapes denote bipolar outflow candidates and outflow candidates that have only a red/blue lobe, respectively. The stars indicate OB stars. The physical scale bar is reported in the bottom right corner of the panel. The rectangles denote IARs.}
\label{Fig:the Perseus arm}
\end{figure}

Figure \ref{Fig:the Perseus arm} shows the spatial distribution of the outflow candidates in the Perseus arm. The OB stars \citep[][and references therein]{XB2018} are marked by stars in Figures \ref{Fig:compare to W5} and \ref{Fig:the Perseus arm} -- \ref{Fig:inter2}.\footnote{We map OB stars in the Perseus arm if their distances $d \in (1.62, 2.97]$ kpc, in the Local arm if $d \in (0.3, 0.94]$ kpc, in the interarm 1 if $d \in (0.94, 1.62]$ kpc, and in the interarm 2 if $d > 2.97$ kpc, but no one in the Outer arm for their large uncertainty of distance.} Most outflow candidates were located at the rectangles marked in Figure \ref{Fig:the Perseus arm}, where the red one indicates the W5 complex that was magnified in Figure \ref{Fig:compare to W5}. The detailed distribution of the outflow candidates in these IARs is discussed in the following subsections.

\subsubsection{W5 Complex} \label{subsection:W5}

\citet{GBW2011} divided the W5 complex into eight sub-regions to analyze the outflows' properties (see figure 3 and section 4 in their paper).\footnote{The eight sub-regions were SH201, AFGL4029, W5 ridge, southern pillars, W5 southeast, W5 southwest, W5 west/IC 1848 and W5 NW which were marked by S201, AFGL4029, LWCas, W5S, W5SE, W5SW, W5W and W5NW in Figure 3 in \citet{GBW2011}, respectively.} It is convenient to describe CO (1 $\rightarrow$ 0) outflow candidates in Figure \ref{Fig:compare to W5} (the eight individual regions are not marked in the figure to avoid complicating the map) based on these sub-regions.

\textbf{Outflow candidates in S201}. This region contains outflow candidates 247, 248, 250 (a high-mass one) and 253. Here the star-forming process was not affected by the HII region Sh 2-201 or by radiation-driven shocks from the nearest W5 O-stars \citep[]{GBW2011}.

\textbf{Outflow candidates in LWCas}. A ridge separated the  W5 complex into two HII region bubbles \citep[i.e., HII regions W5-E and W5-W;][]{DZA2012}. This region contained outflow candidates 214, 215, and 217 -- 222. The consistence of the CO (1 $\rightarrow$ 0) and (3 $\rightarrow$ 2) outflow candidates \citep[20 from][and 215 from this work]{GBW2011} went a step further to confirm this outflow. This region thus provided candidates for radiation-driven implosion and for investigating the relationship between HII regions and a new generation of star formation. As stated by \citet{GBW2011}, this region could serve as an example to explore the transition from molecular to atomic gas under the influence of ionizing radiation regions.

\textbf{Outflow candidates in W5S (containing outflow candidates 212 and 213)}. Three cometary clouds in W5S have been pushed in different directions by the HII region IC 1396 \citep{GBW2011, HZR2013}. Protostars driving the candidate outflows in these cometary clouds were probably triggered by the radiation-driven implosion mechanism \citep[e.g.,][]{GBW2011}.

\textbf{Outflow candidates in W5SW}. Outflow candidate 180 was associated with an isolated clump that showed little evidence of interaction with the HII region \citep[similar to outflow 10 from][]{GBW2011}.

\textbf{Outflow candidates in W5NW}. This region contained outflow candidates 163, 166, 169, 171, 172, 174, 178 and 181, and therefore contained actively forming stars. Similar to the conclusion of \citet{GBW2011}, this region has not been directly impacted by W5 O-stars.

\textbf{Outflow candidates in other regions in the W5 complex}. Outflow candidates 221, 223, 226 and 229 surrounded the east HII region bubble relative to the W5 ridge \citep[i.e., HII region W5-E,][]{DZA2012} which was excited by the double or multiple star HD 18326 \citep[see also the star marked in Figure \ref{Fig:compare to W5},][]{CPO2011, DZA2012, GBV2018} whose spectral type was O6.5V+O9/B0V \citep{SMM2014}.
This implied that these candidates might be affected by the HII region W5-E.

\subsubsection{W4 Complex}

\begin{figure}[!ht]
\centering
\includegraphics[height=0.9\textwidth]{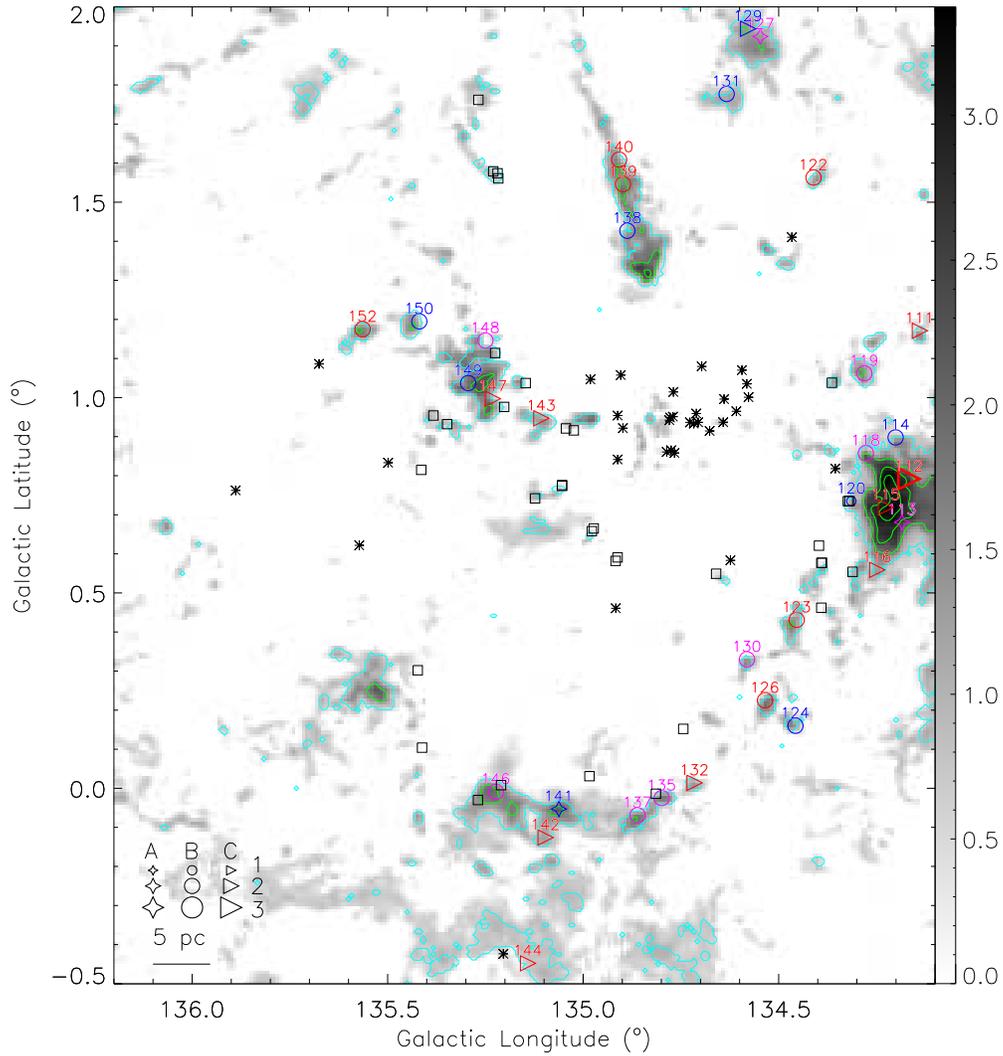}
\caption{Outflow candidate distribution in W4 complex. The description refers to Figure \ref{Fig:the Perseus arm}. The blue/red numbers are the indexes of the outflow candidates that have only a blue/red lobe, and the magenta numbers mark the bipolar outflow candidates. The squares present elephant trunk-like structures in which the star formation is going on possibly due to triggering effect of expanding W4 bubble \citep{PSP2019}.}
\label{Fig:W4}
\end{figure}

The expanding superbubble/HII region W4 was suggested as being driven by the open cluster IC 1805 \citep[centered at the W4 complex;][]{BJM1999, LJ2009}. A total of 126 stars were intrinsic to the open cluster, where the presence of numerous massive stars has been confirmed \citep[about 40 from spectral types O4 to B2, see the OB stars in Figures \ref{Fig:the Perseus arm} -- \ref{Fig:W4}, and][]{SH1999, R2003, LJ2009}. However, this region showed very few CO clouds, which were mainly concentrated to the east of this open cluster as well as in a shell-like structure with a radius of $\sim$ 35 pc \citep[see Figure \ref{Fig:W4} and][]{LJ2009}.

Overall, CO outflow activities in the W4 complex are much fewer than that in the W3 (see below) and W5 complexes. One reason may be that the CO gas was collapsed by UV photons from the OB association IC 1805. Outflow candidates 138 -- 140 were located in a cometary cloud in the north of this region. The HII region W4 was probably responsible for impacting or even triggering the formation of the driving protostars of these three outflow candidates. In the southern region, CO emission in a thin shell (see above) with a width of $\lesssim$ 1 pc was concentrated on the north side (i.e., toward the HII region). The formation of the driving protostars of the ten outflow candidates located at this thin shell were probably impacted by the interaction between the HII regions and the surrounding clouds. Outflow candidates 143 and 147 -- 149 were close to elephant-trunk-like structures (extracted from homogeneous infrared data-sets obtained by the 2MASS, GLIMPSE, MIPS and WISE surveys) in which the star formation therein was possibly due to a triggering effect caused by the expanding W4 bubble \citep{PSP2019}. It was difficult to understand the outflow candidates in the eastern part of the HII region (i.e., 150 and 152). They were weak and could be radiation-driven flows.

\subsubsection{W3 Complex}

The W3 complex is a smaller relative to the W4 and W5 complexes, and is on the western edge of the W4 shell \citep[see Figure \ref{Fig:the Perseus arm} and][]{OWK2005}. W3 North, W3 Main, W3(OH), W3(H$_2$O) and AFGL 333 consist of a ridge that forms a boundary between the W3 and W4 complexes \citep{OWK2005, RMP2011}.

\begin{figure}[!ht]
\centering
\includegraphics[height=0.9\textwidth,angle=-90]{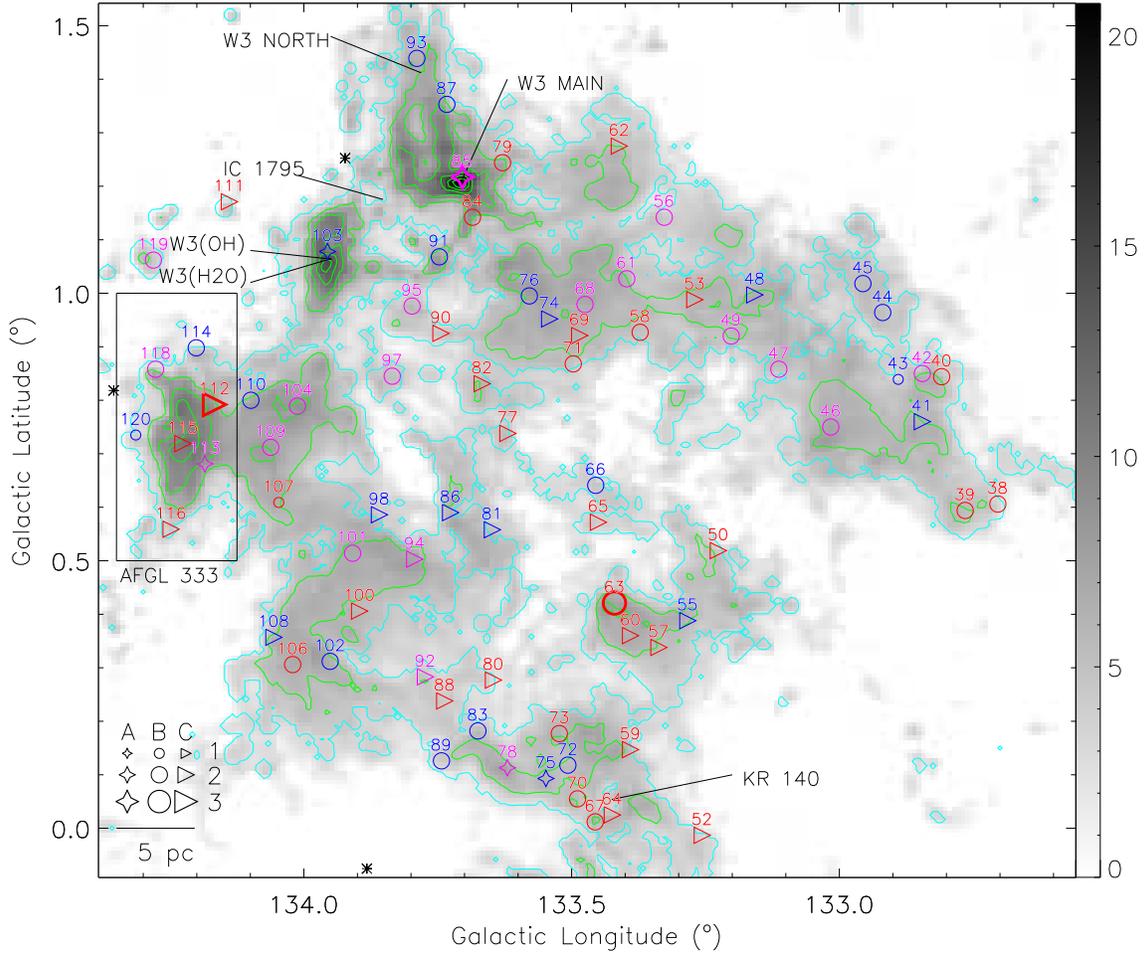}
\caption{Outflow candidate distribution in W3 complex. The description refers to Figure \ref{Fig:the Perseus arm}. The blue/red numbers are the indexes of the outflow candidates that have only a blue/red lobe, and the magenta numbers mark the bipolar outflow candidates.}
\label{Fig:W3}
\end{figure}

Overall, the W3 complex showed much more active star formation in the view of outflow activities relative to the W5 and W4 complexes, i.e., the outflow column density \citep[see e.g.,][]{LLX2018} in the W3 complex ($\sim$ 0.021 pc$^{-2}$) was larger than that in the W5 complex ($\sim$ 0.016 pc$^{-2}$) and in the W4 complex ($\lesssim$ 0.006 pc$^{-2}$). Most of the high-mass outflow candidates were located in the IARs, such as outflow candidates 112 in AFGL 333 and 85 in W3 Main. Outflow candidates for each IARs are described as follows:

In AFGL 333, outflow candidate 115 was associated with an outflow identified from the CO (2 $\rightarrow$ 1) line and an H$_2$O maser \citep[see][]{NSC2017}. They also suggested that star formation in the AFGL 333 region proceeded without significant external triggers.

Hosting a bright O6.5V star, an O9.7I star and two B stars \citep{M1989, OWK2005}, the IC 1795 cluster ionized a diffuse HII region G133.71+1.21 in the W3 complex \citep{WWB1983, RBH2011}. This region contained several sites of high- and low-mass star formations: W3(OH), W3 Main, and W3 North \citep{OWK2005, RBH2011}.

The W3(OH) complex, which contained W3(OH) itself and W3(H$_2$O), was a well-studied high-mass star forming region \citep{QSW2016}. The only outflow candidate, 103, was associated with CO (2 $\rightarrow$ 1) outflows \citep{ZRR2011} and HCN outflows detected by \citet{QSW2016}. A sketch of the dynamical state of the W3(OH) star forming complex can be found in Figure 8 of \citet{QSW2016}.

W3 Main is an ideal region to simultaneously investigate massive star formation at different evolutionary stages \citep{WBZ2012}. High-mass outflow candidate 85 was located in the midst of outflows W3 SMS1 and W3 SMS2 \citep[see figures 11 and 13 in][respectively]{WBZ2012}. Triggered star formation may be responsible for this outflow, because evidence for interactions between the molecular cloud and the HII regions has been found in the W3 Main complex \citep[such as the ultracompact HII regions W3 C and W3 F; see][and references therein]{WBZ2012}.

The bright HII region W3 North (G133.8+1.4) was less well studied than those mentioned above. An O6 star was detected at ($02^\mathrm{h}26^\mathrm{m}49.62^\mathrm{s}, +62\degr15\arcmin35.0\arcsec$)(J2000) via Chandra observations \citep{FT2008}. Because of its small size of $\sim 5\arcmin$ \citep{QRB2006}, it was difficult to see a shell-like structure even if it has one. The nearest outflow candidate, 93, is $\sim 2\arcmin$ to the north of this HII region. This outflow candidate may be impacted by the HII region W3 North.

The HII region KR 140, powered by an O star Ves 735 \citep{KBM1999}, has been investigated in multi-wavelength sub-mm (include 450 and 850 $\mu$m) studies \citep{BKM2000, KMJ2001}. \citet{KMJ2001} detected numerous sub-mm dust cores located at the interface between the ionized and molecular gas, likely implying that the star formation was impacted by the expansion of the HII region. Outflow candidates 59, 64, 70, and 72 were located in a shell-like structure with a diameter of $\sim$ 5 pc that surrounded HII region KR 140 \citep[consistent with the size of this HII region, see][]{KAK2008}. They shared a similar distribution of sub-mm dust cores seen by \citet{KMJ2001}, which indicated that these outflow candidates were probably impacted by the expansion of the HII region.

The outflow candidates in the other regions were well distributed in the shell-like structure of the superbubble/HII region W3 with a radius of $\sim$ 13 pc for the inner side. They were likely to be impacted by the HII region W3. It was noticeable that outflow candidates were distributed nearly uniformly across the field. Such distribution was probably real at least in current angular resolution, because such distribution was obeyed by both B-rated and C-rated outflow candidates (there were only five A-rated outflow candidates) but did not appear in the W4 (Figure \ref{Fig:W4}) and W5 (Figure \ref{Fig:compare to W5}) cloud complex and in the entire Perseus arm regions (Figure \ref{Fig:the Perseus arm}).

\subsubsection{Other Regions in the Perseus Arm}

Some outflow candidates also surrounded OB stars in other regions in the Perseus arm. For example, to the south of the W5 complex, with a projected distance to the boundary of the W5 complex of $\sim$ 42 pc, eight outflow candidates were located in a shell-like cloud (radius is $\sim$ 10 pc) which surrounded the B2III star ALS 7530 \citep[centered at $(l, b) = (138\degr.394, -0\degr.834)$ at a distance of $\sim 2088$ pc,][]{R1978, R2003, G2016}. These outflow candidates were likely impacted by the OB stars.

A small number of outflow candidates were located far from OB stars, indicating that they were untriggered (spontaneous). Overall, the studied regions in the Perseus arm, especially the W3 complex, contained massive stars at various evolutionary stages \citep[e.g.,][]{TGC1997}. This region, especially in the eastern HDL in the W3 complex that neighbors the W4 complex (see the eastern part of Figure \ref{Fig:W3}), contained the most active star-forming sites with signatures of massive star formation in a triggered environment \citep[][]{OWK2005}.

\subsection{The Local Arm}

\begin{figure}[!ht]
\centering
\includegraphics[width=0.77\textwidth]{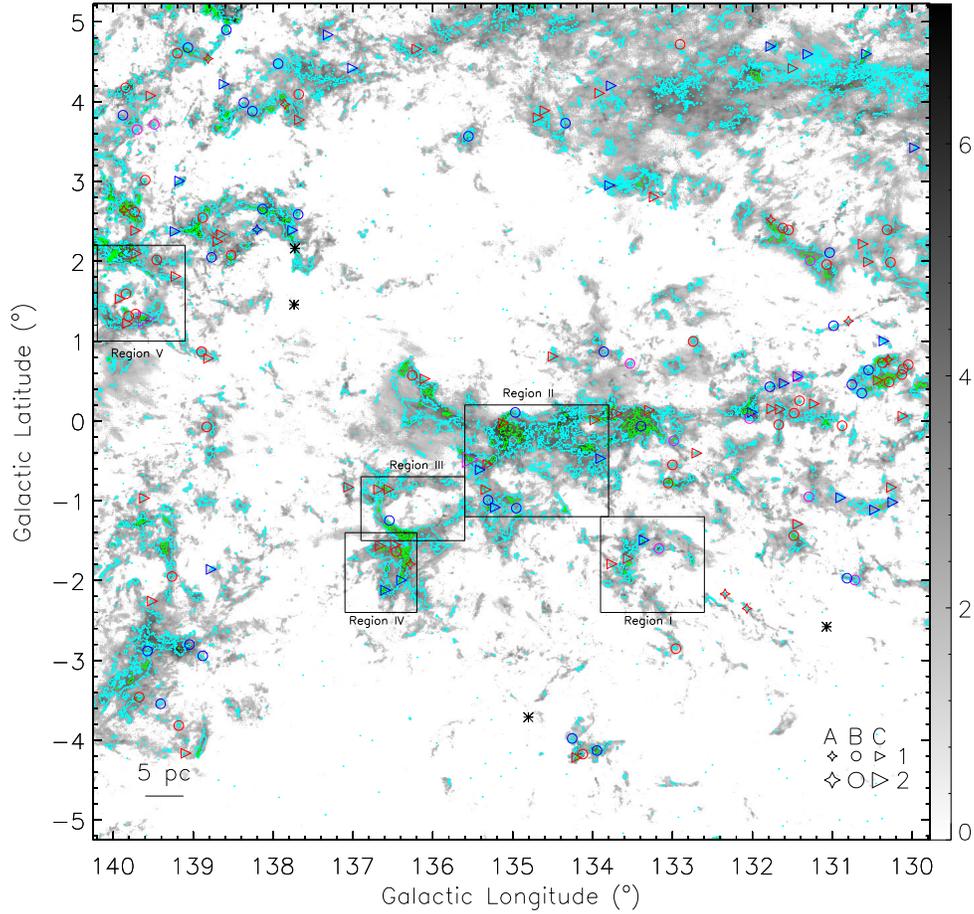}
\caption{Outflow candidate distribution in the Local arm. The background gray-scale map is the integrated intensity map of \co in the range of [-20, 7] km s$^{-1}$, the gray value is the square root of the integrated intensity, and the color bar is in units of (K km s$^{-1}$)$^{1/2}$. The green contours are the integrated intensity map of $^{13}$CO in the same velocity range as the $^{12}$CO map. They start at 5$\sigma$ with $\sigma=0.55$ K km s$^{-1}$, and the contour interval is 20\% of the difference between the peak intensity and 5$\sigma$. The cyan contours denote \xco emission boundaries, where the main beam brightness temperatures are larger than 3 $\times$ RMS in at least three successive channels. The markers to describe quality level and classification of outflow candidates is placed in the bottom right corner of the panel, where 2 and 1 denote intermediate- and low-mass outflow candidates (see the criterion in Section \ref{subsection:parameters}), respectively. The magenta, red/blue colors of the shapes denote bipolar outflow candidates and outflow candidates that have only a red/blue lobe, respectively. The stars indicate OB stars. The physical scale bar is reported in the bottom left corner of the panel. The rectangles denote IARs.}
\label{Fig:loacl arm}
\end{figure}

\begin{figure}[!ht]
\centering
\includegraphics[width=0.77\textwidth]{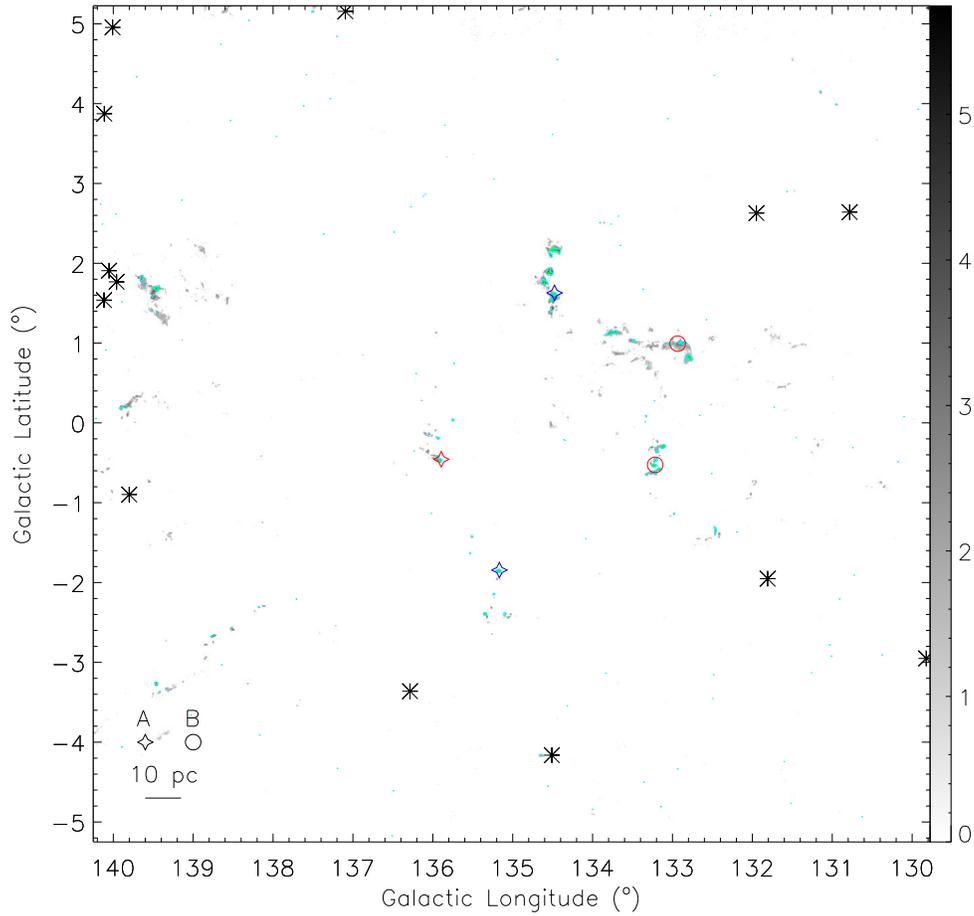}
\caption{Outflow candidate distribution in the interarm 1. The background gray-scale map is the integrated intensity map of \co in the range of [-30, -20] km s$^{-1}$, the gray value is the square root of the integrated intensity, and the color bar is in units of (K km s$^{-1}$)$^{1/2}$. The green contours are the integrated intensity map of $^{13}$CO in the same velocity range as the $^{12}$CO map. They start at 5$\sigma$ with $\sigma=0.33$ K km s$^{-1}$, and the contour interval is 20\% of the difference between the peak intensity and 5$\sigma$. The cyan contour denotes \xco emission boundaries, where the main beam brightness temperatures are larger than 3 $\times$ RMS in at least three successive channels. The markers to describe quality level of outflow candidates is placed in the bottom left corner of the panel. The red/blue color of the shapes denote outflow candidates that have only a red/blue lobe, respectively. All the outflow candidates are low-mass ones (see the Section \ref{subsection:parameters}). The stars indicate OB stars. The physical scale bar is reported in the bottom left corner of the panel.}
\label{Fig:inter1}
\end{figure}

\begin{figure}[!ht]
\centering
\includegraphics[width=0.77\textwidth]{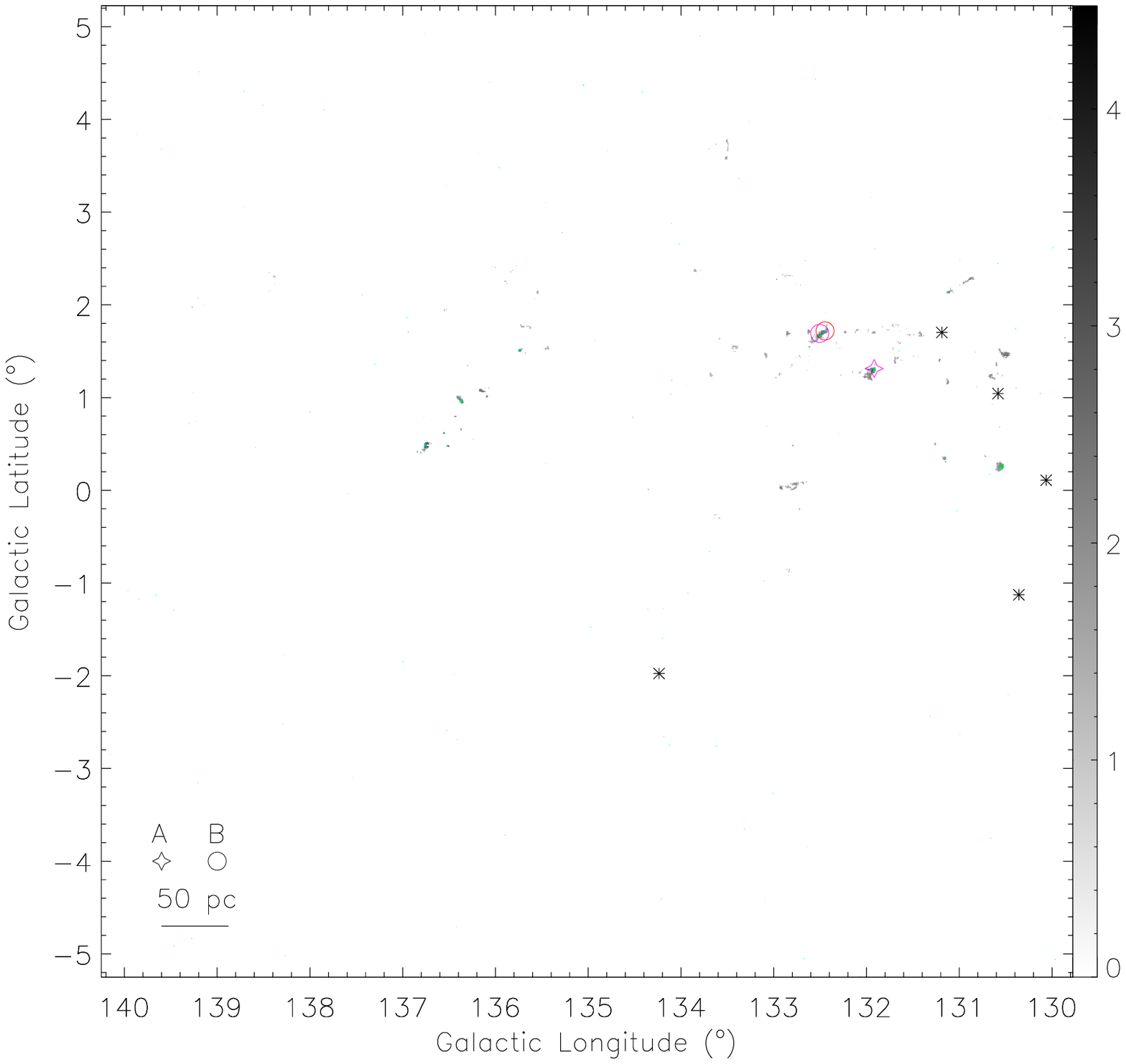}
\caption{Outflow candidate distribution in the interarm 2. The background gray-scale map is the integrated intensity map of \co in the range of [-68, -62] km s$^{-1}$, the gray value is the square root of the integrated intensity, and the color bar is in units of (K km s$^{-1}$)$^{1/2}$. The green contours are the integrated intensity map of $^{13}$COin the same velocity range as the $^{12}$CO map. They start at 5$\sigma$ with $\sigma=0.25$ K km s$^{-1}$, and the contour interval is 20\% of the difference between the peak intensity and 5$\sigma$. The cyan contours denote \xco emission boundaries, where the main beam brightness temperatures are larger than 3 $\times$ RMS in at least three successive channels. The markers to describe quality level of outflow candidates is placed in the bottom left corner of the panel. The magenta and red colors of the shapes denote bipolar outflow candidates and outflow candidates that have only a red lobe, respectively. The stars indicate OB stars. The physical scale bar is reported in the bottom left corner of the panel.}
\label{Fig:inter2}
\end{figure}

\begin{figure}[!ht]
\centering
\includegraphics[width=0.77\textwidth]{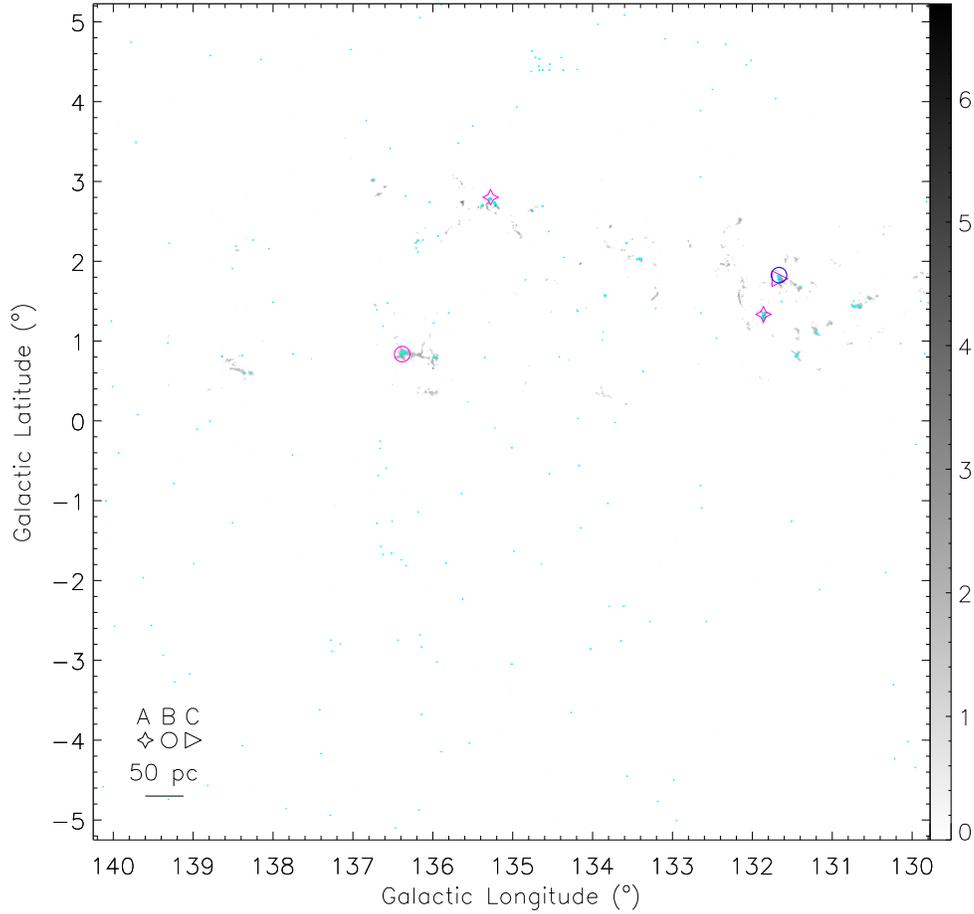}
\caption{Outflow candidate distribution in the Outer arm. The background gray-scale map is the integrated intensity map of \co in the range of [-88, -68] km s$^{-1}$, the gray value is the square root of the integrated intensity, and the color bar is in units of (K km s$^{-1}$)$^{1/2}$. The green contours are the integrated intensity map of $^{13}$CO in the same velocity range as the $^{12}$CO map. They start at 5$\sigma$ with $\sigma=0.46$ K km s$^{-1}$, and the contour interval is 20\% of the difference between the peak intensity and 5$\sigma$. The cyan contours denote \xco emission boundaries, where the main beam brightness temperatures are larger than 3 $\times$ RMS in at least three successive channels. The markers to describe quality level of outflow candidates is placed in the bottom left corner of the panel. The magenta and blue colors of the shapes denote bipolar outflow candidates and outflow candidates that have only a blue lobe, respectively. The physical scale bar is reported in the bottom left corner of the panel.}
\label{Fig:outer}
\end{figure}

Some outflow candidates, including the the only two intermediate-mass ones, were located in shell-like regions (IARs, see the five rectangles marked in Figure \ref{Fig:loacl arm}).
Four outflow candidates were located in a shell-like region (radius of the shell is $\sim 3$ pc) which surrounded an open cluster, Cl Stock 2 with an age of 300 Myr \citep[$l$ = 133.\degr33, $b$ = 1.\degr69,][]{KPS2013, SKP2014, SKP2015}. Surrounding the shell-like structure in the other four IARs (mark by rectangles and labelled as regions II -- V) with radii of approximately 5, 4, 3, and 4 pc, we detected 10 (include an intermediate-mass one), 3, 6 and 8 (include an intermediate-mass one) outflow candidates, respectively. Eight weak OB stars were found to be close to the center of these four IARs in SIMBAD\footnote{see detail in \url{http://simbad.u-strasbg.fr/simbad/sim-fcoo}.}:

\begin{enumerate}
\item For region II they were B9V star TYC 3699-871-1 \citep{HFM2000, CSD2003, G2016}, at a distance of $\sim$ 700 pc and located at ($l, b$) $\sim$ (134$\degr$.8, -0$\degr$.6), and B8/B9 star GSC 03699-00325 \citep{MRM2001, CSD2003, G2016} with a distance of $\sim$ 600 pc located at ($l, b$) $\sim$ (135$\degr$.2, -0$\degr$.6);
\item For region III they were B1 star GSC 0369-02027 \citep{MRM2001, CSD2003, G2016} located at ($l, b$) $\sim$ (136$\degr$.5, -1$\degr$.1) and B9V star HD 15979 \citep{NKA1995, CSD2003, G2016} located at ($l, b$) $\sim$ (136$\degr$.1, -1$\degr$.1) with a distance of $\sim$ 500 pc;
\item For region IV they were B9V star HD 16025 located at ($l, b$) $\sim$ (136$\degr$.5, -2$\degr$.0) and B9 III star HD 16494 located at ($l, b$) $\sim$ (137$\degr$.1, -2$\degr$.0) with a distance of $\sim$ 500 pc \citep{NKA1995, CSD2003, G2016};
\item And for region V they were B9 star TYC 3714-344-1 \citep{HFM2000, CSD2003, G2016} located at ($l, b$) $\sim$ (139$\degr$.7, 1$\degr$.4) with a distance of 800 pc and B8 II-III star V$^{\ast}$ V368 \citep{KKP1971, MRM2001, G2016} located at ($l, b$) $\sim$ (139$\degr$.8, 1$\degr$.7) with a distance of $\sim$ 900 pc.
\end{enumerate}

For region III, the bubble size of the B1 star is $\sim$ 3 -- 8 pc \citep{CZC2013}, and the typical radius of the HII regions was $\sim$ 5 pc \citep{QRB2006}. They are approximated to the radius of the four shell-like structures. If the association among the stars and the shell-like structures is confirmed by future studies, these five IARs could be candidates to study the impact on star formation imposed by massive or intermediate-mass stars.

\subsection{Other Regions} \label{subsection:other regions}

CO clouds in the Outer arm are much rarer than those in the Perseus arm and the Local arm, but were more crowded than those in interarm 1 and interarm 2. The outflow candidates were distributed in these rare CO clouds in the Outer arm, interarm 1 and interarm 2.

In interarm 1, all the outflow candidates were low-mass. The southernmost outflow candidate (index is 450) was located in a relatively isolated cloud, which might be a great example to investigate star formation. The other four outflow candidates were located in filamentary structures that had little mixture to ambient gas.

Because their distances were $\gtrsim$ 4 kpc, the outflow candidates in interarm 2 and the Outer arm were highly clustered (see Section \ref{subsection:parameters}), and their physical parameters are not reported here. We detected two candidate (bulk) outflows\footnote{We denoted the outflow candidates which were highly clustered as candidate (bulk) outflows to differentiate from other outflow candidates.} (the north ones) in a filamentary-like cloud with a length of dozens of pc in interarm 2.

All the candidate (bulk) outflows in the Outer arm were located in filamentary structures with lengths of dozens of pc. The northernmost and the southernmost filamentary structures, especially the former, showed a heavy head or tail, which was evidence for interactions with external forces such as those driven by high- or intermediate-mass stars. For example, the ultracompact HII region IRAS 02395+6244 with ($l, b$) = (135$\degr$.278, 2$\degr$.797) and $V_\mathrm{lsr}$ = -71.5 km s$^{-1}$ \citep{BNM1996, YZW2018}, was probably an indicator of such a driven source.

\section{Discussion}

\subsection{Star Formation Activity}

CO (1 $\rightarrow$ 0) outflow candidates were used to evaluate star formation in this work. Because a protostar outflow is a common phenomenon for star formation in the stage of Class 0 and I \citep[e.g.,][]{B1996, B2016}, we will compare outflow candidates with class 0/I sources in the following discussion.

\citet{KAG2008} detected 171 Class I sources in the W5 complex and \citet{RMP2011} detected 184 Class 0/I and 560 Class 0/I$^{\ast}$ (highly embedded YSOs) sources in W3 Main/(OH), KR 140 and AFGL 333, using Spitzer photometry. Outflows are nearly always associated with Class 0/I objects in nearby star-forming regions \citep[e.g.,][]{HFR2007, CRS2010}. The low number of detected outflow candidates (284 in the Perseus arm, a region that contains W5, W3 Main/(OH), KR 140 and AFGL 333) can be explained as follows.

First, CO outflows were not detected in some Class I sources due to the removal or dissociation of CO gas by the protostar, such as in the HH34 complex \citep[see][]{BD1994, GBW2011, B2016}. Optical or infrared jets can be used to test the association. Second, many outflow candidates were clusters of outflows rather than individual ones. For instance, there were a number of WISE spots located within the contours of the outflow candidate 152. Third, some outflows that exhibited faint emission, low velocity, complex environment, or small size could be missed.

We also detected some outflow candidates that did not have any WISE associations. These outflow candidates were probably driven by first hydrostatic cores, very young protostellar or proto-brown dwarfs, or driven by more evolved protostars with low accretion rates \citep{DVA2014, FPB2018}. These sources may also be fake outflows. Because the Local arm contained a higher proportion of quantity level ``C'' candidates relative to other four regions, we might attribute the high percentage of ``N'' classifications (i.e. no WISE associations) for the Local arm in Table \ref{Fig:statistic} to fake outflows or outflows with the abovementioned driven sources \citep[for examples of outflows driven by candidates with the youngest protostars or first hydrostatic cores see][]{FPB2018}. That is to say, outflow candidates with both ``C'' and ``N'' classifications( $\sim$ 20\%, i.e., 33/162) were more likely to be fake outflows or outflows with the aforementioned driven sources.

\subsection{Discussion of Triggering}\label{discuss:cccc}

The majority of the star formation in the W3/4/5 complex were likely impacted by their corresponding HII regions, although they may not be directly triggered by these HII regions.
In the Local arm, star formation in the IARs marked in Figure \ref{Fig:loacl arm} may be under the impact of massive or intermediate-mass stars. The outflow candidates in interarms 1 and 2 were far from OB stars, indicating that the star formation in these regions was probably untriggered (spontaneous). The OB stars were not shown in the Outer arm due to the large uncertainties in their calculated distances. Nevertheless, we still found that the formation of the driving protostars of the candidate (bulk) outflow (i.e., 458) in the Outer arm could be induced, triggered, or at least impacted by the HII region IRAS 02395+6244 (the projected distance between them was $\sim$ 0.4 pc, see Section \ref{subsection:other regions}). Infrared data, such as in the 5.8, 8.0, 12 and 24 $\mu$m bands \citep{WPC2008, DSA2010, YZW2018}, are helpful when studying HII regions in detail and are therefore advantageous for further investigations regarding induced or triggered star formation (traced by CO (1 $\rightarrow$ 0) outflow candidates) by HII regions.

\section{Summaries}

We conducted a large-scale survey of outflow features toward the W3/4/5 complex region (a total of $\sim$ 110 deg$^2$) using \co and \xco molecular lines. A set of semi-automated IDL scripts based on longitude-latitude-velocity space was used to search for and evaluate outflows over a large area. We detected 459 outflow candidates, of which 284 were in the Perseus arm, 162 in the Local arm, 5 in the Outer arm and the rest (i.e., 8) in interarm regions. Many of the identified outflow candidates were probably multi-outflow candidates resulting from the limited resolution, especially in regions at distances $\gtrsim$ 4 kpc.

As a summary, star formation was mainly concentrated to the Galactic spiral arms. The Perseus arm revealed more activities of stellar formation than those in the Local arm and other regions. There were still a few cases of star formation occurring in the interarm regions, and they were good examples to study relatively isolated star formation. The W3/4/5 complex in the Perseus arm, especially in the eastern HDL where the W3 complex neighbors the W4 complex, presented intense star formation activities. Most of the star formation in these complexes were likely impacted by the HII regions surrounding them.

\acknowledgments
This work is part of the Milky Way Image Scroll Painting (MWISP) project, which is based on observations made with the PMO 13.7 m telescope at Delingha. We would like to thank all the staff members of Qinghai Radio Observing Station at Delingha for their help during the observations. We would like to thank the referee for reviewing the paper carefully and the constructive comments that improves this manuscript. This work was sponsored by the MOST under Grand No. 2017YFA0402701,
the NSFC Grand NO. 11873019, 11673066, 11773077 and 11503033, and the Key Laboratory for Radio Astronomy, CAS.

\facility{PMO 13.7m}

\software{STARLINK/CUPID/FELLWALKER: \citep{BRJ2007, B2015},
GILDAS/CLASS: \citep{P2005}, IDLAstro: (\url{https://idlastro.gsfc.nasa.gov/}), idlcoyote: (\url{http://www.idlcoyote.com/})}

\end{document}